\documentclass[5p,times]{elsarticle}

\usepackage{amssymb,amsmath,xcolor}
\usepackage{soul}

\journal{Physics of the Earth and Planetary Interiors}

\biboptions{authoryear}

\newcommand{\black}[1]{{\color{black}#1}}

\begin{document}

\begin{frontmatter}

\title{Low inertia reversing geodynamos}                      

\author[1]{Chris A. Jones\corref{cor1}}
\cortext[cor1]{Corresponding author}
\ead{c.a.jones@maths.leeds.ac.uk}
\affiliation[1]{organization={Department of Applied Mathematics, University of Leeds},
                city={Leeds},
                postcode={LS2 9JT}, 
                country={United Kingdom}}

\author[2]{Yue-Kin Tsang}
\ead{yue-kin.tsang@newcastle.ac.uk}
\affiliation[2]{organization={School of Mathematics, Statistics and Physics, Newcastle University},
                city={Newcastle upon Tyne},
                postcode={NE1 7RU}, 
                country={United Kingdom}}

\begin{abstract}
Convection driven geodynamo models in rotating spherical geometry have regimes in which reversals occur. However, reversing dynamo models are usually found in regimes where the kinetic and magnetic energy is comparable, so that inertia is playing a significant role in the core dynamics. In the Earth’s core, the Rossby number is very small, and the magnetic energy is much larger than the kinetic energy.  Here we investigate dynamo models in the strong field regime, where magnetic forces have a significant effect on convection. In the core, the strong field is achieved by having the magnetic Prandtl number $Pm$ small, but the Ekman number $E$ extremely small. In simulations, very small $E$ is not possible, but the strong field regime can be reached by increasing $Pm$. However, if $Pm$ is raised while the fluid Prandtl number $Pr$ is fixed at unity, the most common choice, the P\'eclet number becomes small, so that the linear terms in the heat (or composition) equation dominate, which is also far from Earth-like behaviour. Here we increase $Pr$ and Pm together, so that nonlinearity is important in the heat equation and the dynamo is strong field. We find that Earth-like reversals are possible at numerically achievable parameter values, and the simulations have Earth-like magnetic fields away from the times at which it reverses. The magnetic energy is much greater than the kinetic energy except close to the reversal times.    
\end{abstract}


\end{frontmatter}

\setstcolor{red}

\section{Introduction}
The first spherical convection-driven geodynamo model to exhibit a reversal similar to those revealed in the paleomagnetic record was that of \cite{GlatzmaierNature1995}. This simulation used nominal large values of Prandtl number $Pr=5000$ and magnetic Prandtl number $Pm=500$ \citep{GlatzmaierPepi1995}, but as hyperdiffusion was used the effective values of these parameters might have been different. \cite{Kutzner2002} also found dynamo models with occasional reversals without the use of hyperdiffusion, with $Pr=1$, $Pm=3$. Following this work, reversing dynamo models at $Pr=1$ have been explored further, see e.g. \cite{Driscoll2009b} and \cite{Sprain}. However, in all these $Pr=1$ dynamo reversal models, the magnetic and kinetic energies are comparable, and inertia may play an important role in inducing the reversals. In the Earth's core, the magnetic energy is much larger than the kinetic energy, and inertia 
is unimportant, except on very short length scales. Indeed, 
\cite{Schaeffer2017}  say this about reversing dynamo models \lq Furthermore, we emphasize that, to our knowledge, no simulation with strong field (i.e. magnetic energy larger than kinetic energy) has ever exhibited polarity reversals\rq .

\cite{Sarson1999} and \cite{Sarson2000} found reversing dynamos when the inertial term was removed, and the momentum equation is solved as a diagnostic equation for \black{the velocity} ${\bf u}$ rather than a prognostic equation timestepping ${\bf u}$. At the time, computers were such that the resolution was limited, and removing inertia entirely leaves the question of when it becomes negligible unanswered. So  these two papers suggested that reversing dynamos might exist when the kinetic energy is small compared to the magnetic energy, the low inertia regime, but were not conclusive. The main purpose of this paper is to exploit the large increase in computer power since 1999 to further investigate, with well-resolved models, if reversing strong-field dynamos exist at low inertia, and if they have \lq Earth-like\rq \, magnetic morphology
when the dipole is strong, as it is currently.  

\cite{Glatzmaier1999} found reversals in their dynamo model, which were affected by the inhomogeneous boundary condition at the  \black{core--mantle boundary (CMB)}, and this could either slow down or speed up the reversal rate depending on whether the heat flux taken out of the core was enhanced at the poles or the equator. \cite{Olson2010} looked again at heterogeneous boundaries, using the moderate $E$ models of \cite{Driscoll2009} rather than the Glatzmaier-Roberts dynamo model, finding similar results to those of 
\cite{Glatzmaier1999}. The same may well be true of the models studied here, but we have not explored this yet, preferring to study the simpler homogeneous case first. 

\cite{Christensen2006} (see also \cite{Olson2006}) introduced the concept of the local Rossby number with $Ro_{l} \approx 0.1$ as the boundary between dipolar and multipolar dynamos. Earth-like reversing dynamos were found near this boundary. Since then, this boundary region has been taken as the criterion for Earth-like dynamos to exist. The global Rossby number can be written in terms of the standard dynamo parameters as 
$Ro=E Rm/Pm$, $E$ being the Ekman number, $Rm$ the magnetic Reynolds number and $Pm$ being the magnetic Prandtl number. For the parameters where reversal simulations are feasible, the local Rossby number is typically between 2 to 8 times larger than $Ro$ itself. It is not practical to reduce $E$ much below $10^{-4}$ if very long runs, needed to investigate whether reversals occur, are required. Unless $Rm$ is over about 50 dynamo action does not occur in spherical dynamo simulations, and it has to be several hundred to get a strong field dynamo needed to give a realistic model. So to reduce inertia, the only practical option is to increase $Pm$ and this is what we do here. However, If $Pm$ is large and $Pr$ is left at unity, the codensity equation is dominated by the diffusive term rather than the codensity advection term. This is certainly not the correct balance in planetary cores, and it has the effect of reducing the fluctuations introduced through the buoyancy. So as well as increasing $Pm$ we also increase $Pr$ so that the ratio, the Roberts number $q=Pm/Pr$ is of order 1 or less. This has the effect of making the P\'eclet number comparable to $Rm$, rather than having a very small P\'eclet number. It might seem odd to study dynamos at large $Pm$ and $Pr$ since in the Earth's core $Pm$ is $O(10^{-6})$ and $Pr$ is small for liquid metals and its value in the compositional driving case is even smaller. However \cite{Dormy2016} showed that because the Ekman number in dynamo models is much larger than its true value, $Pm$ needs to be large to achieve the strong field regime in which Lorentz forces play a significant role. It is more important for the force balances that hold in the core to be respected than it is for $Pm$ and $Pr$ to have their correct values. \cite{Aubert2019} has demonstrated that there is evidence for a one parameter family of solutions which show Earth-like behaviour with $Pm$ gradually reducing with Ekman number ($Pm \sim E^{1/2}$ was suggested) so that very small $Pm$ is reached when $E$ achieves its $O(10^{-15})$ value. Here we propose that for reversing dynamos a scaling in which $Pr$ reduces with $E$ might also be appropriate, leading to a path from large to small $Pr$ along which Earth-like reversal behaviour might be found.

\cite{Driscoll2009} looked at a range of dynamos with $E \sim 10^{-3}$ and explored where reversing dynamos occur, and \cite{Driscoll2009b} considered the variation of reversal frequency with an evolving core model. The local Rossby number criterion highlighted the importance of the balance between inertia and Coriolis forces being appropriate for reversals to occur. This was a somewhat worrying development, because reversals certainly occur in the geomagnetic field even though the inertial forces are very small compared to the Coriolis forces, thus raising the issue of whether the whole picture of the geodynamo being driven by convection influenced by rotation might be incorrect. However, \cite{Sreenivasan2014} argued that although inertia was significant in their $Pr=1$, $Pm=5$ simulations, it was actually the buoyancy force rather than inertia that was breaking down the dipole dominance and allowing the dynamo to reverse. We are building on this picture by letting $Pr$ increase, so that nonlinear advection in the buoyancy equation becomes important.    

\cite{Christensen2010}
gives some parameters which generate models with magnetic fields similar to that at the Earth's \black{CMB}, but which \black{do not} reverse. Some of these models satisfied a particular set of criteria which are satisfied by the current geomagnetic field. These criteria can be used to decide whether it is Earth-like or not, providing  a useful simple test of whether the generated field looks like the geomagnetic field at the current time. We take the view that for a reversing geodynamo model to be described as Earth-like, there must be some times when the field satisfies the \cite{Christensen2010} criteria.    

\cite{Sheyko2016} found reversing (almost periodic) dynamos at very low $E$ which had the form of dynamo waves. Inertia is important in these models, because it drives a zonal flow, and the zonal flow stimulates the $\alpha-\omega$ dynamo waves.
Because of the dynamo-wave origin, the reversals were rather regular, almost periodic, which is quite unlike the behaviour of geomagnetic reversals, which is more like that of a Poisson process \citep{Cox1968,Constable2000}. \cite{Sprain} gave an assessment of how $Pr=1$ dynamos have performed compared to paleomagnetic data, considering whether the reversal behaviour was compatible with the geomagnetic reversal record or not. An important criterion was the amount of time spent during the reversal process compared with the average time between reversals. For the geomagnetic field this ratio is quite small, as the average interval between reversals is around $3 \times 10^5$ years, while the duration of a reversal is generally less than $10^4$ years.

There have also been a number of stochastic models of geomagnetic reversals, of varying degrees of sophistication. In these \black{models, }the fluctuations are added by an explicit random term pushing the solution off a steady dipole. An early one is \cite{Schmitt2001}, which is based on the idea of a particle trapped in a potential well with two symmetric minima and a local maximum at the origin. The particle is randomly forced, and with small forcing remains near one of the minima. An exceptionally large fluctuation can get the particle over the central maximum corresponding to a reversal. By tuning the forcing, the statistics of geomagnetic reversals can be reproduced. More sophisticated models are \cite{Benzi2010}, \cite{Meduri2016} and \cite{Carbone2020}, (see also references within). \cite{Molina-Cardin2021} propose a stochastic model that reproduces the temporal asymmetry of geomagnetic reversals, with slower decaying rates before the reversal and faster growing rates after it. As we see below, numerical dynamo models also have random fluctuations, and occasionally reverse like the stochastic models. However, the stochastic models are essentially just a description of the observed behaviour, and they are not able to give much insight into the physics of why reversals occur, and whether they are likely to occur on other planets or are unique to the geodynamo.

In section 2 we discuss the model equations and the diagnostic quantities we use to analyse the results. In section 3 we give the results for models driven by a fixed heat flux through the core, and in section 4 we discuss models based \black{upon} compositional convection, where a buoyancy flux at the inner core boundary is mixed into the core interior. Section 5 contains the discussion and conclusions.

\section{Equations solved}

\subsection{Dimensional Boussinesq equations with codensity $C'$}

The equations on which our dynamo simulations are  based are
\begin{eqnarray}
 \rho \left( \frac{{\textrm D} {\bf u}}{{\textrm D} t} + 2{\bf \Omega} \times {\bf u} \right)  &=&  -\nabla p' + \rho g  C' {\bf \hat r} \nonumber \\
&& + \frac{1}{ \mu}(\nabla\times {\bf B})\times {\bf B} + \rho \nu \nabla^2 {\bf u},
 \label{eq:dim-mom}\\
\frac{\partial {\bf B}}{\partial t} &=& \nabla\times({\bf  u} \times {\bf B}) + \eta \nabla^2 {\bf B}, \label{eq:dim-ind}\\[0.1cm]
%
\frac{{\textrm D} C'}{{\textrm D} t} &=&  \kappa \nabla^2 C' - \frac{S}{\rho}, \label{eq:dim-codens1}
 \\[0.1cm]
\nabla\cdot {\bf u} &=& 0, \qquad \nabla \cdot {\bf B} = 0. \label{eq:dim-div}
\end{eqnarray}
Here ${\bf u}$ is the fluid velocity, $\bf \Omega$ the rotation vector of the Earth, $p'$ the pressure, $\rho$ the density, $g$ the gravity, which increases linearly with $r$, consistent with the almost constant density assumption, $C'$ is the codensity, the mass fraction of light element in the compositional case, and proportional to temperature for thermal convection. 
$\mu = 4 \pi \times 10^{-7} \, Hm^{-1}$ \black{is} the permeability of free space, ${\bf B}$ is the magnetic field, $\nu$ the kinematic viscosity, $\eta$ the magnetic diffusivity, $\kappa$ the mass diffusion coefficient of the light material in the compositional case and the thermal diffusivity in the thermal convection case, and $S$ is the sink (or source) term.
$\kappa$ is much lower in the compositional case than it is in the thermal case. 

These are the standard Boussinesq dynamo equations for a spherical, convecting, rotating shell in the codensity formulation. We consider two cases, motivated by previous work on dynamo reversals \citep{Sprain}. The first 
corresponds to thermal convection driven by a fixed flux of heat through the inner core boundary (ICB), and leaving at the CMB. There is no heat source in this model, so the time-averaged total heat flux entering the ICB equals the time-averaged heat flux passing through the CMB. Following \cite{Sprain} we denote this case as FFFF. The second model we consider is motivated by compositional convection. Here a flux of light element is released at the ICB as the outer core solidifies, and this light material is mixed into the outer core. There is therefore a sink of buoyancy in the interior, which balances the input light element flux released at the ICB as outer core material freezes, so on average there is no flux of light material through the CMB. We denote this case of compositional convection boundaries by CCB.


\subsection{Dimensionless Boussinesq equations}

The unit of length is $d=r_o-r_i$, $r_o$ being the outer core radius and $r_i$ being the inner core radius, and the unit of time is $d^2/\eta$, the magnetic diffusion time. The unit of magnetic field is $(\mu \rho \Omega \eta)^{1/2}$. The unit of $C'$ in the thermal convection case is $\kappa \alpha \beta d / \eta$, where $-\beta$ is the temperature gradient at $r=r_o$  and $\alpha$ is the coefficient of thermal expansion. In the compositional case, the unit of $C'$ is $S d^2/\rho \eta$. The unit of pressure $p'$ is $\rho \Omega \eta$, the unit of density is $\rho$
and $\bf \hat z$ is the unit vector parallel to the rotation axis. The dimensionless equations are, in the form solved in the code,

\begin{eqnarray}
\frac{{\textrm D} {\bf u}}{{\textrm D} t} + 2\frac{Pm}{E} {\bf \hat z} \times {\bf u}  &=&  -\frac{Pm}{E} \nabla p' + \bigg(\frac{Ra Pm^2}{Pr}\bigg) C' {\bf r} \nonumber \\
&& + \frac{Pm}{E} (\nabla\times {\bf B})\times {\bf B} + Pm \nabla^2 {\bf u}, \label{eq:dimless-mom}
 \\
\frac{\partial {\bf B}}{\partial t} &=& \nabla\times({\bf u} \times {\bf B}) +  \nabla^2 {\bf B}, \label{eq:dimless-ind}\\[0.1cm]
\frac{Pr}{Pm}\frac{{\textrm D} C'}{{\textrm D} t} &=&  \nabla^2 C', \quad {\rm FFFF} \ {\rm case} , \label{eq:dimlesscodens1}
\\[0.1cm]
\frac{Pr}{Pm}\frac{{\textrm D} C'}{{\textrm D} t} &=&  \nabla^2 C' - \frac{Pr}{Pm}, \quad {\rm CCB} \ {\rm case}, \label{eq:dimlesscodens2}
\\[0.1cm]
\nabla\cdot {\bf u} &=& 0, \qquad
\nabla \cdot {\bf B} = 0. \label{eq:dimless-div}
\end{eqnarray}
The dimensionless parameters are the Ekman number, the Prandtl number and the magnetic Prandtl number,
\begin{eqnarray}
 \quad E = \frac{\nu}{\Omega d^2}, \quad Pr=\frac{\nu}{\kappa}, \quad Pm = \frac{\nu}{\eta},
\label{eq:parmdefs}
\end{eqnarray}
and the Rayleigh number in the fixed flux thermal convection model and the compositional convection model are respectively
\begin{eqnarray}
Ra=\frac{g_0 \alpha \beta  d^5}{r_o \eta \nu}, \qquad   Ra=\frac{g_0 S d^6}{r_o \rho \eta \kappa \nu}.
\label{eq:Ra_defs}
\end{eqnarray}
Here $g_0$ is gravity at the CMB ($r=r_0$), and in the thermal case $C' =  \alpha T$,
where $\alpha$ is the coefficient of thermal expansion and $T$ is the temperature. Note there is variation in the definition of these
parameters in the literature, particularly the Rayleigh number, so when comparing results using different codes it is important to ensure that equivalent
parameter values are used. The dimensionless velocity is in magnetic Reynolds number units, so that the root mean square value of $|{\bf u}|$, the dimensionless velocity over the outer core, is $Rm$. The magnetic field is in Elsasser number units, so the root mean square value of 
$|{\bf B}|$ is the Elsasser number. The magnetic energy is output in units of $\rho \eta^2 / d^2$, as is the kinetic energy, so magnetic energy $ME$ and the kinetic energy $KE$ are
\begin{eqnarray}
ME&=&\int_V \frac{Pm}{2E}  \black{|{\bf B}|^2} \, r^2 \sin \theta \, d\phi \, d\theta \, dr, \nonumber \\
KE&=&\int_V \frac{1}{2}  \black{|{\bf u}|^2}\, r^2 \sin \theta \,  d\phi \, d\theta \, dr.
\label{eq:energydefs}
\end{eqnarray}
The dimensionless volume $V= 4 \pi (r_o^3-r_i^3)/3 d^3=14.5988$ for the standard radius ratio $\lambda = r_i/r_o =0.35$, so the magnetic Reynolds number, $Rm$, the P\'eclet number $Pe$, the global Rossby number $Ro$, and the Elsasser number $\Lambda$, which all play a significant role in this problem, are 
\begin{eqnarray}
Rm&=&\left( \frac{2 {\rm KE}}{V} \right)^{1/2}, \quad \ Pe=\frac{RmPr}{Pm}, \nonumber \\
Ro&=&\frac{ E Rm}{Pm}, \qquad \Lambda = \left( \frac{E}{Pm} \frac{2\rm ME}{V} \right)^{1/2}.  \label{eq:output_parms}
\end{eqnarray}
We also define the local Rossby number $Ro_{l}$ \citep{Christensen2006},
\begin{equation}
Ro_{l} = \frac{\bar l_u}{\pi} Ro, \quad {\bar l_u} = \frac{\sum_l l ({\bf u}_l \cdot {\bf u}_l)}{2 KE}, \label{eq:local_Rossby}
\end{equation}
where the sum is over the spherical harmonic contributions and  \black{${\bf u}_l$} is the component of the velocity of degree $l$, so that
$\bar l_u$ is a weighted mean spherical harmonic degree, representing a \lq typical' wavenumber of the velocity field. 

\subsection{Boundary conditions}

The boundary conditions are no-slip, so ${\bf u}=0$ at the CMB and ICB, and electrically insulating, so ${\bf B}$ matches onto a current-free potential field inside the inner core and outside the CMB in all runs. Two types of codensity boundary conditions were used. The runs in tables 1 and 2 used fixed flux FFFF boundary conditions,  so the codensity fluctuations satisfy $\partial C'/\partial r =0 $ at the ICB and CMB for all spherical harmonics except for $l = m = 0$, and there is no source, so the total flux through the ICB equals that through the CMB on average. For the $l = m = 0$ component we set the 
dimensionless codensity gradient to $-Pr /\lambda^2 Pm$ on the ICB and then the zero source means that on average (but not
instantaneously) the CMB temperature gradient is $-Pr/Pm$. Additionally we impose the $l=m=0$ component of the codensity to be
zero on the CMB, to fix the arbitrary constant in $C'$ that would be present if flux conditions are used exclusively.  

The runs in table 3 used boundary conditions mimicking compositional convection, the CCB boundaries, where  
$\partial C'/\partial r =0 $ at the ICB and CMB for every $l,m$ spherical harmonic coefficient except $l=m=0$. For the component $l=m=0$, the compositional flux on the ICB equals the total absorbed by the sink, so  
\begin{eqnarray}
 \frac{\partial C'}{\partial r} = -\frac{Pr}{Pm} \frac{1-\lambda^3}{3\lambda^2(1-\lambda)}.
\label{eq:CCB}
\end{eqnarray}
The other boundary condition is that the $l=m=0$ component of $C'$ is zero at the CMB as with the FFFF conditions.  

\subsection{Conversion to dimensional units}

The core radii are $r_o=3.485 \times 10^6$\,m, $r_i=1.220 \times 10^6$\,m, so the unit of length, $d=2.265 \times 10^6$\,m. \cite{Pozzo2012} give the electrical conductivity as $1.1 \times 10^6$\,S\,m$^{-1}$. Since $\eta = 1/\sigma \mu$, $\eta=0.723$\,m$^2$\,s$^{-1}$. Note this value is lower than the value of $2$\,m$^2$\,s$^{-1}$
commonly used before 2012, and some believe the older value of 2\black{\,m$^2$\,s$^{-1}$} may be correct. However, we adopt $\eta=0.723$\,m$^2$\,s$^{-1}$ here. The diffusion time $d^2/\eta =7.096 \times 10^{12}$\,s, which is 225,000 years. 

To get the magnetic field unit, we need $\rho$. The mass of the whole core is $1.941 \times 10^{24}$\,kg, so the mean density of the core is 10,950\,kg\,m$^{-3}$, and since the inner core is small we take this as the mean density of the outer core. $\Omega=7.29 \times 10^{-5}$\,s$^{-1}$, so the unit of magnetic field is
\begin{eqnarray}
\left( \Omega \mu \rho \eta \right)^{1/2} = 0.85{\rm mT}. 
\label{eq:Bunit}
\end{eqnarray}

A typical estimate for the velocity in the core is $U_0 = 5 \times 10^{-4}$\,m\,s$^{-1}$, so the magnetic Reynolds number $U_0 d / \eta \approx 1600$. This is high for numerical simulations, and we mostly work in the range $Rm \sim 200-1000$. Note that the effectively larger value of $\eta$ than the true value may have the effect of reducing the effective diffusive timescale in our simulations, so that although a simulation of one diffusion time is nominally 225,000 years it might really only represent a somewhat shorter time.

\subsubsection{Dipole moment}

 At the Earth's surface,
\begin{eqnarray}
B_r = \frac{\mu}{4 \pi} \frac{2m_0 \cos \theta}{r_e^3}, \label{eq:Br_surf}
\end{eqnarray}
where $m_0$ is the dipole moment in A\,m$^2$, and $B_r$ is the radial field due to the axial dipole component only, and $r_e$ is the radius of the Earth. The magnetic field at the Earth's surface from the axial dipole Gauss coefficient $g_{10}$ is
\begin{eqnarray}
B_r = 2 g_{10} \cos \theta,  \label{eq:g10_def}
\end{eqnarray}
so the dipole moment can be expressed in terms of $g_{10}$ as
\begin{eqnarray}
m_0 = \frac{4 \pi}{\mu}  {r_e}^3 g_{10} = 7.614 \times 10^{22} {\rm A\,m}^2   \label{eq:m0_value}
\end{eqnarray}
setting $r_e=6.371 \times 10^6$\,m and $g_{10}=29,442$\,nT which is the value in IGRF for the year 2015. Note that if the geomagnetic field was purely dipolar, the radial field at the pole at the Earth's surface would be $5.89 \times 10^{-5}$\,T, not far off its actual value, and  $3.6 \times 10^{-4}$\,T at the pole of the CMB. Actually the CMB field is only roughly dipolar, but this is a crude estimate of the CMB field strength. In the results below, since the CMB is the upper boundary of our computation, we show the dimensionless value of $g_{10}$ evaluated at the CMB rather than the Earth's surface, where $g_{10}|_{cmb}$ = $(r_e^3/r_o^3)g_{10}|_{surf}$.
This gives $g_{10}|_{cmb} = 1.8 \times 10^{-4}$ T in the year 2015, and using eqn.\,(\ref{eq:Bunit}) 
the dimensionless $g_{10}|_{cmb}=0.2116$ then. This result does depend on the somewhat uncertainly known value of $\eta$; with $\eta = 2$\,m$^2$\,s$^{-1}$, then the dimensionless dipole coefficient was 0.127 in 2015. These values are in the right ball-park looking at the results figures below, which is quite surprising given how far off the dimensionless parameters are.

A further point to bear in mind is that the current geomagnetic field has a strong dipole compared to that found by paleomagnetic data, the current value being approximately twice the average value over the last 160 million years \citep{Juarez1998}.    

\subsection{Alternative form of the dimensionless Boussinesq equations}

An alternative but equivalent set of equations is
\begin{align}
E_m \frac{{\rm D} {\bf u}}{{\rm D} t} + 2 {\bf \hat z} \times {\bf u}  &=  - \nabla p' + {\cal R} C' {\bf r} + (\nabla\times {\bf B})\times {\bf B} + E \nabla^2 {\bf u}, \label{eq:alt-dimless-mom}
\\%
\frac{\partial {\bf B}}{\partial t} &= \nabla\times({\bf u} \times {\bf B}) +  \nabla^2 {\bf B},  \label{eq:alt-dimless-ind}
\end{align}

\begin{equation}
\frac{{\textrm D} C'}{{\textrm D} t} = q \nabla^2 C', \quad {\rm FFFF}; \quad \frac{{\textrm D} C'}{{\textrm D} t} = q \nabla^2 C' - 1,
\quad {\rm CCB};
\label{eq:alt-dimless-codens} 
\end{equation}
\begin{equation}
\nabla\cdot {\bf u} = 0, \qquad
\nabla \cdot {\bf B} = 0.\label{eq:alt-dimless-div}
\end{equation}
Here
\begin{equation}
{\cal R} = Ra E \frac{Pm}{Pr}, \qquad
E_m = \frac{E}{Pm}, \qquad q = \frac{Pm}{Pr}. \label{eq:alt-dimless-parms}
\end{equation}
There is a one-to-one correspondence between $[Ra, E, Pr, Pm]$ and $[{\cal R}, E, E_m, q]$. Here ${\cal R}$ is the rotationally modified Rayleigh number, $E_m$ is the magnetic Ekman number $\eta/ \Omega d^2$, and $q$ is the Roberts number $\kappa/\eta$ (small in the Earth's core).

The advantage of this formulation is that it makes it clearer how balance between the Coriolis, Lorentz and buoyancy forces (more precisely between the curls of these forces, MAC balance)  is to be achieved. Clearly $E_m$ and $E$ must be small. Note that inertia scales like $({\bf u} \cdot \nabla ){\bf u}$, and the dimensionless ${\bf u}$ is comparable with $Rm$ which has to be at least 200 in Earth-like dynamos, so it is $E_m Rm = Ro$, the Rossby number which must be small. Actually, since there is a derivative in the inertial term, and it is the vorticity balance from eqn.\,(\ref{eq:alt-dimless-mom})  that must be respected, $Ro$ must be quite small for inertia to be really negligible.   

\subsection{Criteria for Earth-like reversals}

As noted by \cite{Sprain}, many dynamos have a small dipolar component and reverse frequently. These are commonly known as multipolar dynamos. This is not Earth-like behaviour, as the Earth has a mean reversal time comparable with a \black{magnetic} diffusion time, and has had superchron periods lasting hundreds of magnetic diffusion times. Superchrons may well be caused by processes not included in elementary models such as those discussed here, so we do not insist on the existence of superchrons for an Earth-like reversal model. We classify our models into three types, non-reversing (type N), Earth-like (type E) and multipolar (type M)  
in a way that is broadly similar (but not identical to) \cite{Sprain}. For a dynamo to be of type E, it must reverse, but it must also have periods of at least one magnetic diffusion time in which it stays in one polarity, and there must be at least one such period for both polarities. This excludes frequently reversing multipolar dynamos. We know that the Earth's dynamo has periods when it is strongly dipole dominant, and indeed it is currently in such a state.  
There may have been times when the geomagnetic field was weaker, and reversed more frequently \cite[e.g.][]{Gallet2019}, so occasional
intervals of multipolar behaviour are not necessarily incompatible with the paleomagnetic data.  

\cite{Christensen2010} give a convenient measure of dipole dominance, the ratio of the power of the axial dipole field to the rest of the field up to degree and order 8, 
\begin{eqnarray}
    AD/NAD = P_{10} / \left( P_{11} + \sum_{l=2}^{8} \left( \frac{r_e}{r_o} \right)^{(2l-2)} \sum_{m=0}^{\black{l}} P_{lm} \right)
    \label{eq:ADNAD}
\end{eqnarray}
where $r_e$ is Earth's radius, $r_o$ is the core radius, and
\begin{eqnarray}
    P_{lm} = (l+1) \left( g_{lm}^2 +h_{lm}^2 \right)  \label{eq:Plm_def}
\end{eqnarray}
is the power in a component of degree \black{$l$} and order $m$ at the Earth's surface. This ratio is currently about 1, though it has been 1.5 in the recent past. The dipole is currently quite strong compared to the mean value over the paleomagnetic record. We consider that a field model is acceptable as an Earth-like reversing dynamo
provided that AD/NAD has been greater than 1 in polarity states of opposite signs. We did test the models for equatorial symmetry and zonality \citep{Christensen2010}, but most of the dynamos passed these tests, so they were not very discriminating. The test for flux concentration depends strongly on how the field model is derived from the data (almost all numerical dynamo models have very high flux concentration if run at high resolution) so this was not used as a criterion. To summarise, a dynamo model gets an \lq E' classification if (a) the axial dipole changes sign, (b) there are periods of at least one magnetic diffusion time during which the dynamo does not reverse, and there must be such periods with $g_{10}$ both positive and negative,
(c) the ratio AD/NAD must be greater than unity at some time both when $g_{10}$ positive and when it is negative. For some runs, the dynamo mostly had an axial dipole of one sign, except for one or two excursions where it briefly changed sign but did not establish the opposite polarity regime. We denote these as N*.
If they were run for much longer, these N* models might become Earth-like, but they would be very expensive to study in detail. 

Another simple measure of dipolarity was used, based on the relative strength of the $g_{10}$ coefficient to the mean magnetic field.
\begin{gather}
Dip= \frac{dip_{rms}}{\bar B}, \quad
dip_{rms} = \left( \frac{1}{\tau} \int_0^\tau g^2_{10} \, d \mbox{\black{t}} \right)^{1/2}, \nonumber \\
{\bar B} = \left(\frac{1}{\tau V} \int_0^\tau \int_V \black{|}{\bf B}\black{|}^2 \,  r^2 \sin \theta \, d\phi \, d\theta \, dr \, dt \right)^{1/2},
\label{eq:Dipdef}
\end{gather}
where the integrals are over the whole length of the run and the whole volume of the outer core. Because the field inside the core is stronger than the surface field, a value of $Dip \approx 0.04$ or above gives a field pattern which is Earth-like for some of the time. 

Another useful diagnostic is the dipole tilt angle. The dipole tilt angle $\varphi$ is related to the Gauss coefficients,  as follows:
\begin{align}
\cos\varphi &= \frac{g_{10}}{\sqrt{g_{11}^2 + h_{11}^2 + g_{10}^2}} .
\label{eq:tilt}
\end{align}
The dipole tilt angle provides a simple way to test whether the field is in a dipolar or multipolar regime. In a dipolar regime, $\varphi$ remains close to 0 or 180$^\circ$, but in a multipolar regime $\varphi$ fluctuates wildly.


\section{Results from the dynamos driven from below}

Eqs. (\ref{eq:dimless-mom} - \ref{eq:dimless-div}) were solved numerically using the pseudo-spectral Leeds spherical dynamo code \citep{Willis2007}.
Table~\ref{table:cases1} gives a list of the runs performed with $E=2 \times 10^{-4}$, and table~\ref{table:cases2} the list for $E=10^{-4}$. The code uses a high-order finite difference scheme in the radial direction, with $Nr$ points spaced at the zeroes of a Chebyshev polynomial. $Nr=160$ radial points was found to be sufficient for all the runs reported here.
The variables are expanded in spherical harmonics, and the maximum degree is $Nl$ and the maximum order is $Nm$. For most of these runs $Nl=Nm=128$ was adequate, but for $E= 10^{-4}$ some runs were  checked with $Nl=Nm=192$ to make sure there was no significant difference.  

All runs generated long lasting magnetic fields, so they are all dynamos. These runs use the fixed flux, no-slip boundaries
 with no internal heat source, the FFFF conditions. Originally, it was intended to use the compositional convection boundary conditions only. However, it emerged that Earth-like dynamo reversals with FFFF conditions occurred at lower $Rm$ and slightly higher $E$ than CCB conditions, which meant that it was feasible to explore the parameter space with FFFF boundary conditions, whereas this was not possible for the more numerically demanding CCB case. However, we did some long runs with compositional convection boundaries, see section 4 below. The criterion for deciding on the resolution necessary is that the energy spectra of the spherical harmonic decomposition should be such that all quantities had less than 1\% magnitude at the highest harmonics than they had at the maximum harmonic. In practice, high resolution of the magnetic field is the hardest to achieve (the flow was well-resolved) and this determined the number of spherical harmonics required. 
 
 The tables give the three input parameters, $Pr$, $Ra$ and $q$ ($Pm=qPr$), the modified Rayleigh number ${\cal R}$ and the output parameters magnetic Reynolds number $Rm$, P\'eclet number $Pe=Rm/q$, and global Rossby number $Ro$. Since $\bar l_u$ (see eqn. (\ref{eq:local_Rossby})) is typically between $15$ and $20$ in these runs, the local Rossby number is typically a little less than 0.01,
 much smaller than the 0.1 figure that signalled Earth-like reversing dynamos at $Pr=1$. $\tau$ is the length of the run in the dimensionless unit, so the nominal dimensional time of the run is $\tau \eta/d^2$. Note that all runs were for at least several magnetic diffusion times, and at the large $Pr$ and $Pm$ this corresponds to a great many thermal and viscous diffusion times and at least a thousand convective turnover times. These are very long runs which unfortunately do consume a significant amount of computational resource. The output parameter $Dip$ is given by (\ref{eq:Dipdef}).
In the reversal column in the tables, an N indicates a non-reversing dynamo (no reversal throughout the whole run), an M denotes a multipolar dynamo, frequently reversing, and an E denotes an \lq Earth-like' dynamo that reverses occasionally but has long periods (at least a whole magnetic diffusion time)
when it has a strong dipole moment of constant sign. 
If we adopt the \cite{Pozzo2012} value of $\eta$ the magnetic diffusion time $d^2/\eta$ is 225,000 years and $Rm \approx 1600$. It can be argued that the convective turnover time is the most relevant timescale in the dynamo process, and with the values used in section 2.4, $d/U_0 \approx 140$ years. In our simulations the value of $Rm$ used is less than 1600 for numerical reasons, so our convective turnover time is a factor $1600/Rm$ longer than that of the real Earth. It is therefore possible to argue that one unit of dimensionless time corresponds to only $225,000 Rm/1600$ years, typically around 50,000 years only. 

\begin{figure*}
\centering
\begin{minipage}{0.48\textwidth}
(a)\\
\includegraphics[width=0.95\textwidth]{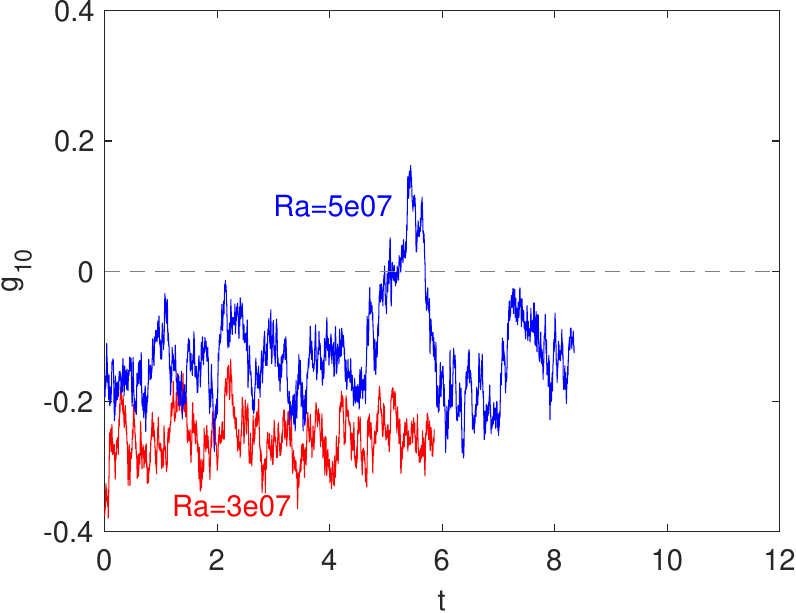}\\
(c)\\
\includegraphics[width=0.95\textwidth]{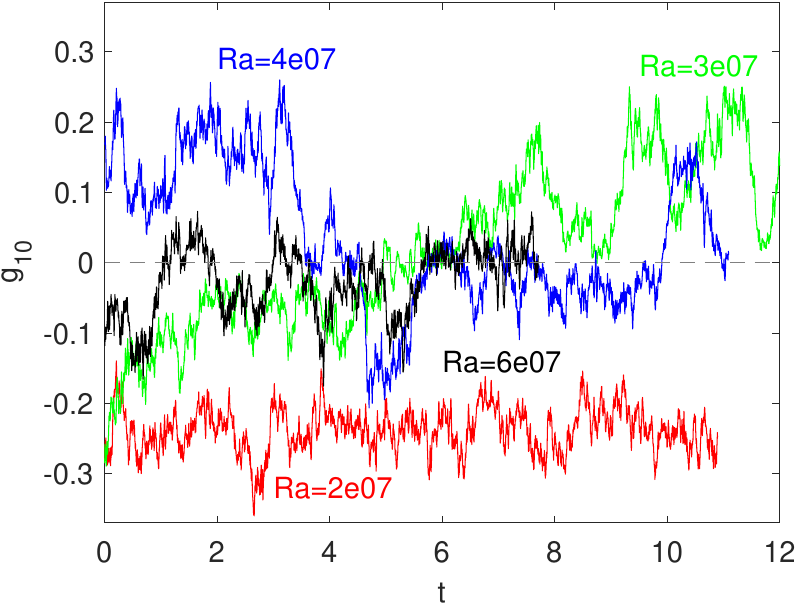} 
\end{minipage}
\begin{minipage}{0.48\textwidth}
(b)\\
\includegraphics[width=0.95\textwidth]{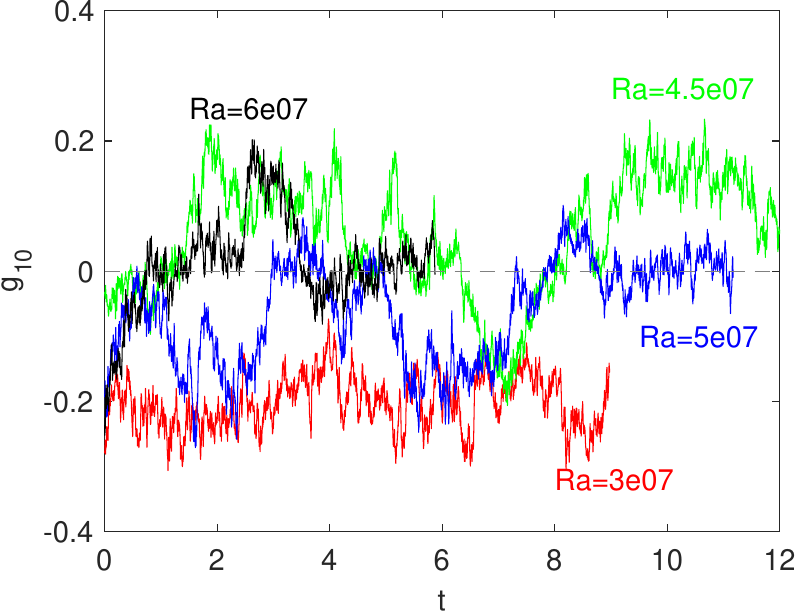}\\
(d)\\
\includegraphics[width=0.95\textwidth]{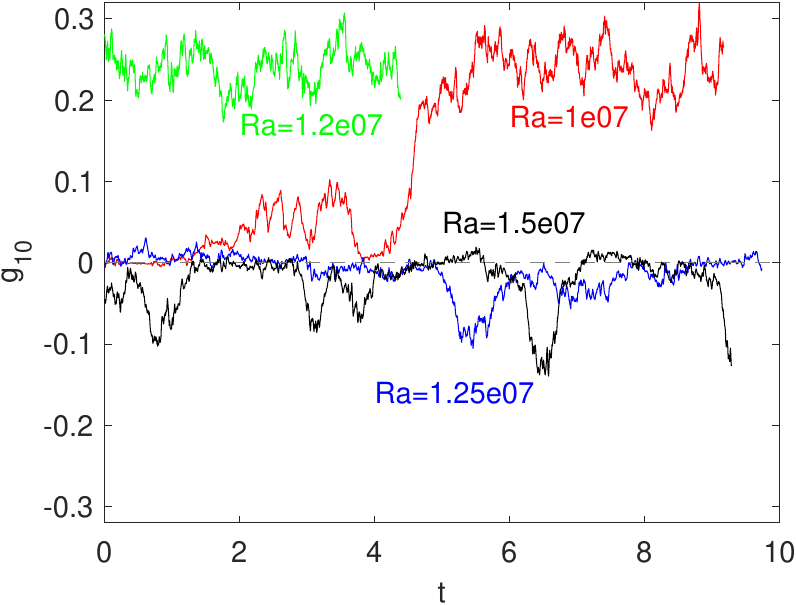} 
\end{minipage}

\caption{Axial dipole coefficient $g_{10}$ at the CMB as a function of magnetic diffusion time $t$ with $E=2 \times 10^{-4}$, $Pm=40$. \ (a) $Pr=30$, $Ra=3 \times 10^7$ and $Ra=5 \times 10^7$, runs B1, B2. \  (b)  $Pr=35$, $Ra=3 \times 10^7$, $Ra=4.5 \times 10^7$, $Ra=5 \times 10^7$, and $Ra=6 \times 10^7$, runs C1, C3, C4, C5.\ (c) $Pr=40$, $Ra=2 \times 10^7$, $Ra=3 \times 10^7$, $Ra=4 \times10^7$ and $Ra=6 \times 10^7$, runs D1, D2, D3, D6. \ (d) $Pr=50$, $Ra = 10^7$, $Ra=1.2 \times 10^7$, $Ra=1.25 \times 10^7$, and $Ra=1.5 \times 10^7$, runs E1, E2, E3, E4. The grey dashed line is where $g_{10}=0$}.
\label{figs:fig1}
\end{figure*}

\begin{figure*}
\centering
\begin{minipage}{0.48\textwidth}
(a)\\
\includegraphics[width=0.95\textwidth]{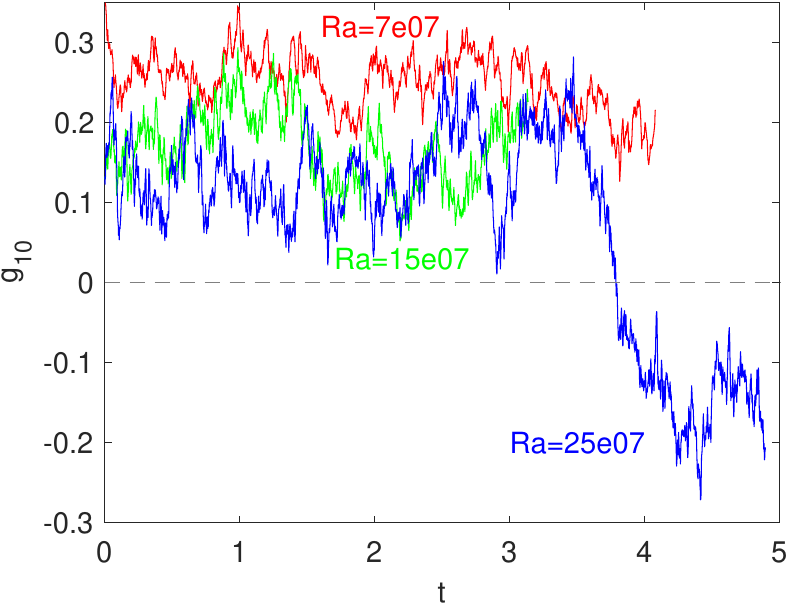}\\
(c)\\
\includegraphics[width=0.95\textwidth]{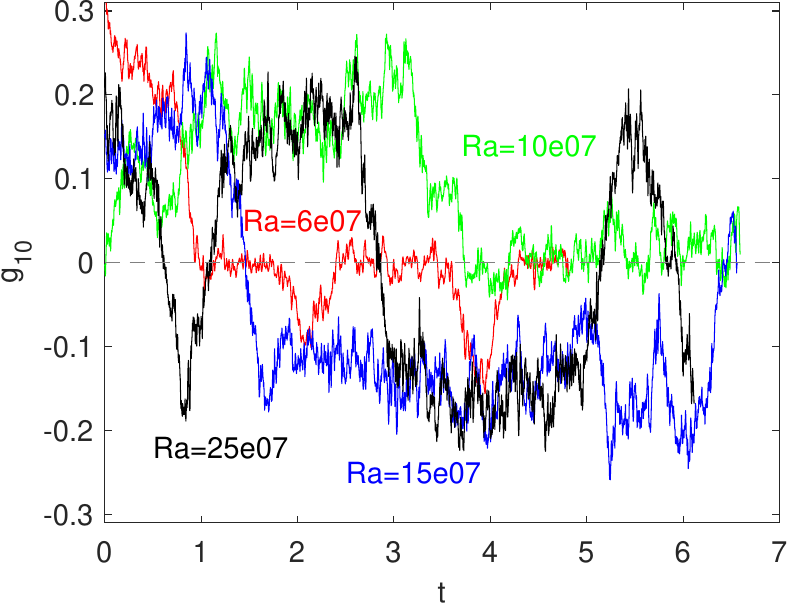} 
\end{minipage}
\begin{minipage}{0.48\textwidth}
(b)\\
\includegraphics[width=0.95\textwidth]{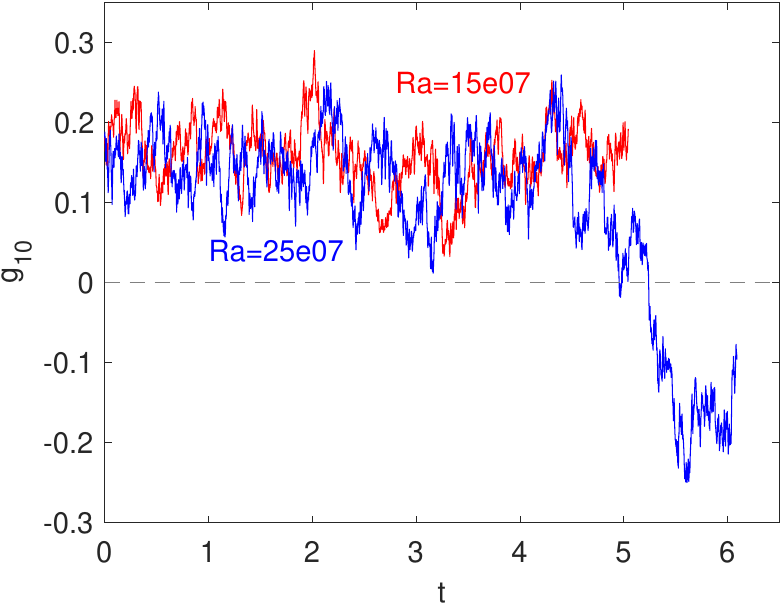}\\
(d)\\
\includegraphics[width=0.95\textwidth]{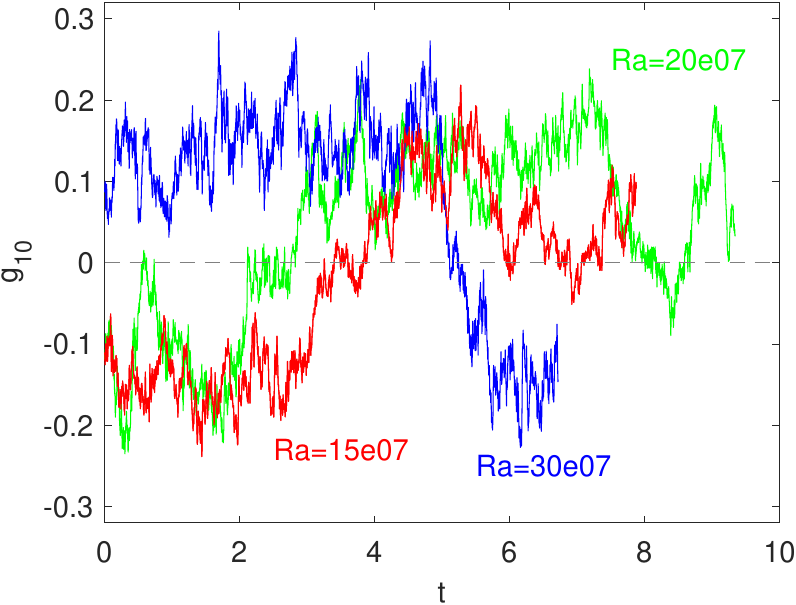} 
\end{minipage}

\caption{Axial dipole coefficient $g_{10}$ at the CMB as a function of magnetic diffusion time $t$ with $E= 10^{-4}$, $Pm=40$. \ (a) $Pr=50$, $Ra=7 \times 10^7$, $Ra=1.5 \times 10^8$ and $Ra=2.5 \times 10^8$, runs H1, H4, H5. \  (b)  $Pr=55$, $Ra=1.5 \times 10^8$ and $Ra=2.5 \times 10^8$, runs
I2, I3.\ (c) $Pr=60$, $Ra=6 \times 10^7$, $Ra= 10^8$, $Ra=1.5 \times10^8$ and $Ra=2.5 \times 10^8$, runs J4, J6, J7, J8. \ (d) $Pr=70$, $Ra = 1.5 \times 10^8$, $Ra=2 \times 10^8$, $Ra= 3 \times 10^8$, runs K1, K2, K3.}
\label{figs:fig2}
\end{figure*}

In Figs.~\ref{figs:fig1} and \ref{figs:fig2}, the axial dipole coefficient $g_{10}$ at the CMB is plotted as a function of time, as a change of sign of $g_{10}$ corresponds to a reversal. 

%
%
%
%

\subsection{$E=2 \times 10^{-4}$ runs}

Six series of runs were performed with $E=2 \times 10^{-4}$, with $Pr$ varying from $Pr=20$ to $Pr=80$, see table 1. All these runs had $Pm=40$,
so the Roberts number $q$ lies in the range $0.5 \leq q \leq 2$. For each value of $Pr$ the Rayleigh number was varied so that the magnetic Reynolds number (which increases with $Ra$) lies in the range $150 < Rm < 500$, large enough to give reversing dynamos but not so large as to be too computationally demanding. 
$Pm=40$ is sufficient to ensure a low enough Rossby number so that inertia does not play a significant role, and makes the magnetic energy much larger than the kinetic energy.  None of our runs have the huge ratio of magnetic to kinetic energy expected in the real Earth, but they are at least in the correct regime as regards energy ratio. At the smallest value of $Pr=20$ (runs A1 and A2) the dynamo \black{did not} reverse. It may be that at higher values of $Rm$ than we could reach there may be reversals, as expected from previous work. Fig. 1a shows two runs at $Pr=30$, B1 and B2, one at $Ra= 3 \times 10^7$ which \black{did not} reverse, and one at $Ra=5 \times 10^7$ which did reverse and had marginally \lq Earth-like' reversal behaviour in that it did reverse occasionally, but it did not have sustained times with both polarities. 

Fig. 1b has runs C1, C3, C4 and C5 at $Pr=35$ so $q$ has been reduced to 1.143 and the P\'eclet number is almost as large as $Rm$. Run C1 at $Ra=3 \times 10^7$ \black{did not} reverse, but C3 at $Ra=4.5 \times 10^7$ is reasonably Earth-like, reversing and with a large dipolar component most of the time. More information about runs C3 and D2 is shown in Fig. \ref{figs:fig4}. Run C4 at $Ra=5 \times 10^7$ is on the border between Earth-like and multipolar. It does have significant periods with a large dipole component, though the positive $g_{10}$ interval was less than a full diffusion time, but there are also long periods where the field is definitely not dipole dominated, and so the field pattern at the CMB would fail the \cite{Christensen2010} criteria for an Earth-like dynamo then. Run C5 at  \black{$Ra=6 \times 10^7$} has occasional periods of dipole dominance, but most of the time it is multipolar. The conclusion is that there is a sweet spot at around $Rm \approx 400$, corresponding to $Pe \approx 370$, where Earth-like behaviour is found. For $Rm$ smaller the fluctuations of the dipole about the mean are too small to get reversals, while for $Rm$ larger, the fluctuations are too large and the dynamo is multipolar and not Earth-like. 

Fig. 1c is for the four D1, D2, D3 and D6 runs at $Pr=40$. Run D1 at $Ra=2 \times 10^7$ never reversed, but D2 at $Ra=3 \times 10^7$ has quite Earth-like reversal behaviour, see also Fig. \ref{figs:fig4}, and it was also run using the MAGIC code (see Fig. \ref{figs:fig10}) to check that the reversal behaviour was robust between two independent numerical codes. Run D3 was on the border between Earth-like and multipolar, reversing but with uncomfortably long times with a rather weak dipolar component, and run D6 is definitely multipolar. So as with the C runs, there is a sweet spot for Earth-like behaviour for the D runs, but it happens at a slightly lower value of $Rm$ between 300-350. \black{The value of $Pe$  is a better guide than $Rm$ to when the field transitions from the non-reversing N state to an Earth-like E state. Run A2 is non-reversing despite having an $Rm=450$, while run D2 with $Rm$ only 315 has Earth-like reversals. Run A2 has a smaller $Pe=225$ than run D2, where $Pe=315$. This suggests that the fluctuation levels, and hence the ability to reverse, are more controlled by $Pe$ than $Rm$. However, the transition from N to E is not simply a matter of $Pe$ crossing a threshold, for run C2 did not reverse while D2 did, despite them having similar $Pe$.}    

Fig. 1d shows results for $Pr=50$ so $q$ is reduced to 0.8,
runs E1, E2, E3, and E4. None of these runs showed Earth-like behaviour, but they did have interesting features. Run E1 at \black{$Ra=1.0 \times 10^7$} had a small $g_{10}$ which stayed small for 4 diffusion times, but it then grew to a larger value, and ended up as a strongly dipolar solution.  Run E2 at $Ra=1.2 \times 10^7$ was started from a strongly dipolar solution and showed no sign of reversing. Run E3, despite having only a very slightly larger $Ra=1.25 \times 10^7$, had a completely different behaviour, almost entirely multipolar, with only a few weakly dipolar stretches. This strongly suggests subcritical behaviour with two different solutions existing. For computational reasons it was not possible to do a full investigation of exactly when the two branches become unstable, but no Earth-like reversal behaviour was found at this (or larger) values of $Pr$ because there is no stable sweet spot between the non-reversing and multipolar solutions. The larger $Ra$ solution E4 was mainly multipolar, as were the higher $Ra$ runs, E5 and E6, and all the F runs at $Pr=80$, $q=0.5$ were multipolar also.  

%
%
%
%
%

%
%
%

\begin{figure*}
\centering
\begin{minipage}{0.48\textwidth}
(a)\\
\includegraphics[width=0.95\textwidth]{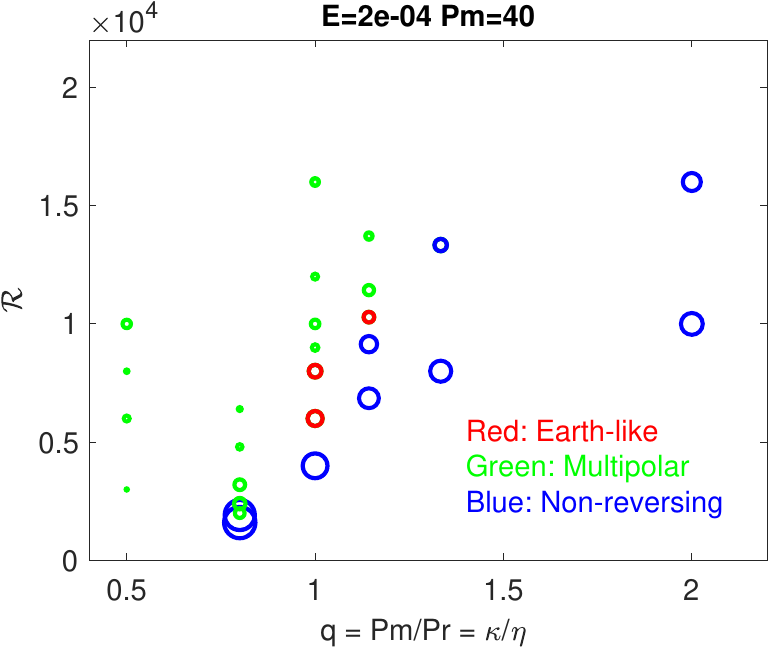} 
\end{minipage}
\begin{minipage}{0.48\textwidth}
(b)\\
\includegraphics[width=0.95\textwidth]{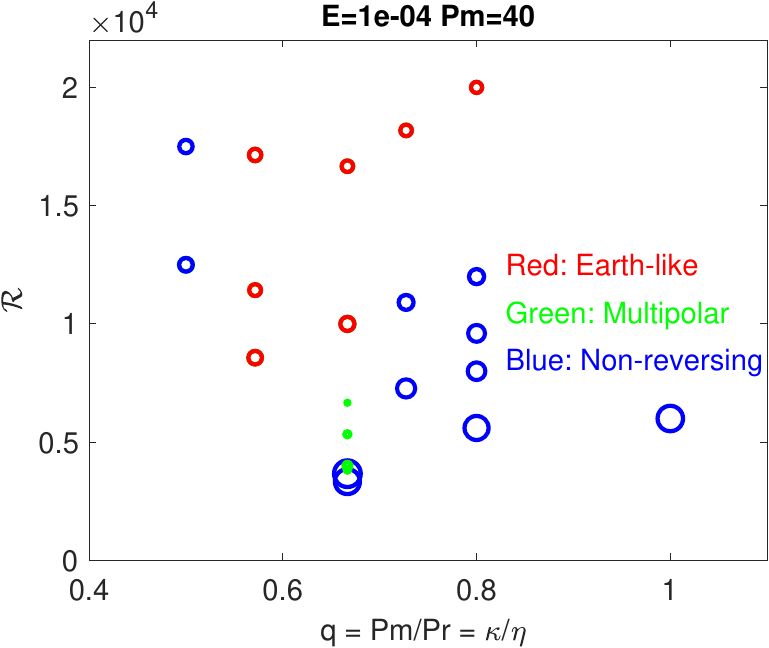} 
\end{minipage}

\caption{Scatter plots of the fixed flux no source dynamos in the $q$-${\cal R}$ plane, non-reversing dynamos are blue circles, multipolar dynamos are green, and
Earth-like dynamos are red. \  (a) $E=2 \times 10^{-4}$. \ (b) 
$E=10^{-4}$. The radii of the plotted circles are proportional to $Dip = dip_{rms} / {\bar B}$, so larger circles correspond to dipole dominated fields, small circles to  dynamos with only a relatively small $g_{10}$ component. }
\label{figs:fig3}
\end{figure*}

\subsection{$E= 10^{-4}$ runs}

With currently available computational resources it is not possible to reduce $E$ very much, but by making a modest reduction to $E=10^{-4}$ we can see whether the reversal behaviour is becoming more Earth-like at lower $E$ or less Earth-like. From table 2 and Fig. 2 we see that Earth-like reversal behaviour was found for all $Pr$ between 50 and 70, so the desired reversal behaviour is more common at lower $E$, though of course the runs are computationally more expensive. However, it is necessary to reduce $q$ to get reversing dynamos at lower $E$. At $E=2 \times 10^{-4}$ Earth-like reversals were found at $q >1$ but at $E=10^{-4}$ runs
at $q=1$ or above were non-reversing for the feasible range of $Rm$. Since the values of $q$ and $E$ are both very low in the Earth's core, it is possible that an asymptotic path leading to Earth-like values may exist. In Fig. 2a at $Pr=50$,  $q=0.8$, we see that the two lower Rayleigh number runs H1 and H4, with $Rm=315$ and $466$ \black{did not} reverse, but run H5 with $Rm=588$ had Earth-like reversals. Similar behaviour was found at $Pr=55$ for the two runs I2 and I3, see Fig.2b. Run I3, which has Earth-like reversals, has a slightly lower $Rm$  than run H5, but a slightly higher $Pe$ which may be why it reverses. 

The situation at $Pr=60$, Fig. 2c is a little more complicated. At $Ra=5 \times 10^7$ where $Rm=253$ only, the dynamo is a non-reversing rather steady dipole dominated 
dynamo. But by $Ra=6 \times 10^7$, run J4, the dipole component collapsed to leave a multipolar dynamo with only occasional excursions to a dipolar state, which is though less dipolar than the current Earth. Increasing $Ra$ to $10^8$, run J6, still gave a multipolar dynamo, but further increasing to $Ra=1.5 \times 10^8$, run J7, led to an Earth-like dynamo, as did the run at $Ra=2.5 \times 10^8$, run J8. This re-establishment of reversing dynamos with strong dipolar states at $E=10^{-4}$ was not found at $E=2 \times 10^{-4}$ and it is why Earth-like dynamos are more common at the lower Ekman number. This behaviour
was also found at $Pr=70$, see Fig.2d runs K1,K2 and K3, which were all Earth-like, though the higher $Ra$ runs had a stronger dipolar component when they were not in the reversal phase. 

Somewhat unexpectedly, further increasing $Pr$ to 80, thus reducing $q$ to 0.5, led to a non-reversing dynamo even with $Rm=537$, though we caution that these $Pr=80$ runs were shorter, as this part of the parameter space is even more computationally demanding. \black{It would be interesting to discover why non-reversing dynamos are found here,
where we expect strong fluctuations because both $Pe$ and $Re$ are large. Possibly the fluctuations here are centred on shorter wavelength modes, and so do not affect the dipole component so much.} Our main conclusion from all these runs is that reversing dynamos with long periods of dipole dominance do occur at the lowest values of $E$ that are numerically feasible. \black{Within the range $0.8 \geq q \geq 0.571$ all dynamos with $Pe > 650$ had Earth-like behaviour.} As with the $E=2 \times 10^{-4}$ runs there is a preferred interval of $q$ values where Earth-like reversals occur at moderate (and therefore numerically accessible) $Rm$. Fortunately, the interval of $q$ seems to broaden out as $E$ is reduced, as well as being centred on a lower value of $q$. 

\subsection{Dynamos in the ${q-\cal R}$ plane}

We can summarise the results from tables 1 and 2, and Figs. 1 and 2, by showing the locations of Earth-like, non-reversing and multipolar dynamos in the $q-{\cal R}$ plane, Figs. 3a and 3b. The dynamos are plotted as circles centred on their $q$ and $\cal R$ values, with radius proportional to $Dip = dip_{rms} / {\bar B}$ which measures how dipolar the field is. Typically the non-reversing (blue) dynamos are the most dipole-dominated and the green multipolar dynamos the least dipolar. The Earth-like (red) dynamos have moderate dipolarity, and their ratio of the axial dipole to the non-axial terms is broadly similar to that of the Earth. The location of the Earth-like reversing dynamos is similar in Figs. 3a and 3b except that it is peaked around a lower $q$ in the lower Ekman number case, and the region is also broader at lower $E$.  At the lower value of $E$ not many multipolar solutions were found, but they may  exist at higher values of $\cal R$ where computation is more difficult. 

%
%
%
%

\begin{figure*}
\centering
\begin{minipage}{0.48\textwidth}
(a)\\
\includegraphics[width=0.95\textwidth]{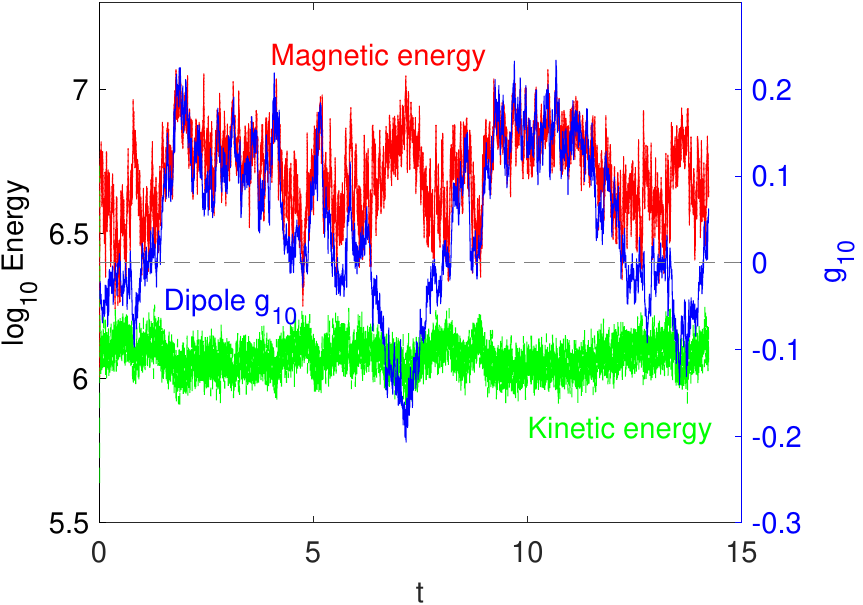}\\
(c)\\
\includegraphics[width=0.95\textwidth]{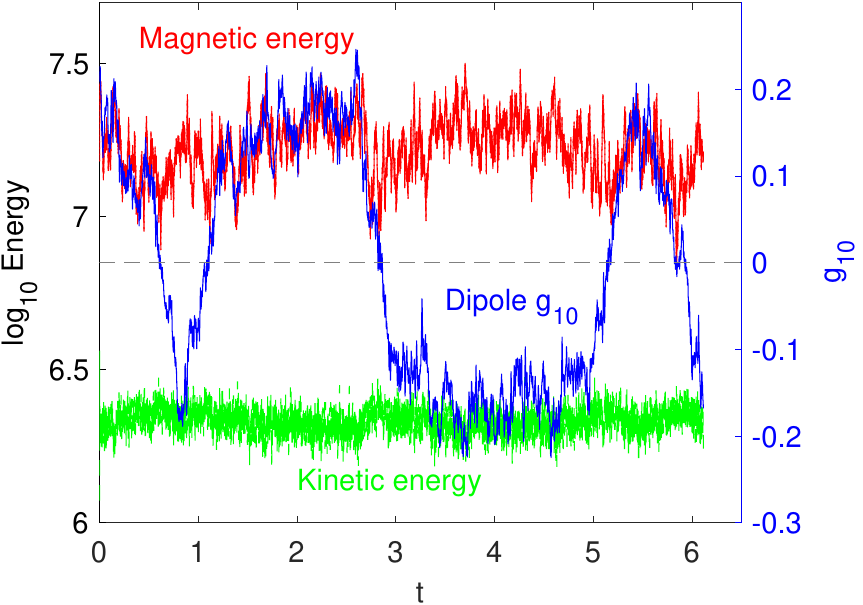} 
\end{minipage}
\begin{minipage}{0.48\textwidth}
(b)\\
\includegraphics[width=0.95\textwidth]{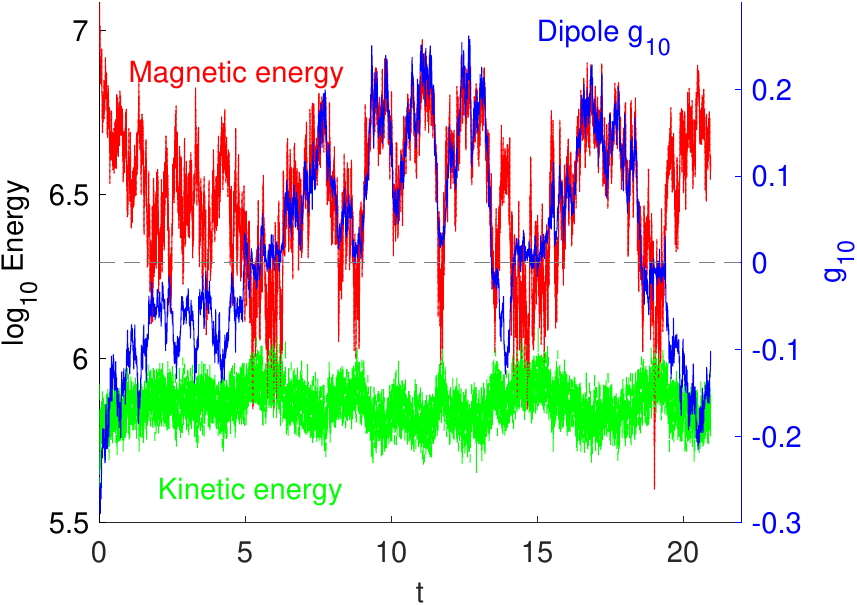}\\
(d)\\
\includegraphics[width=0.95\textwidth]{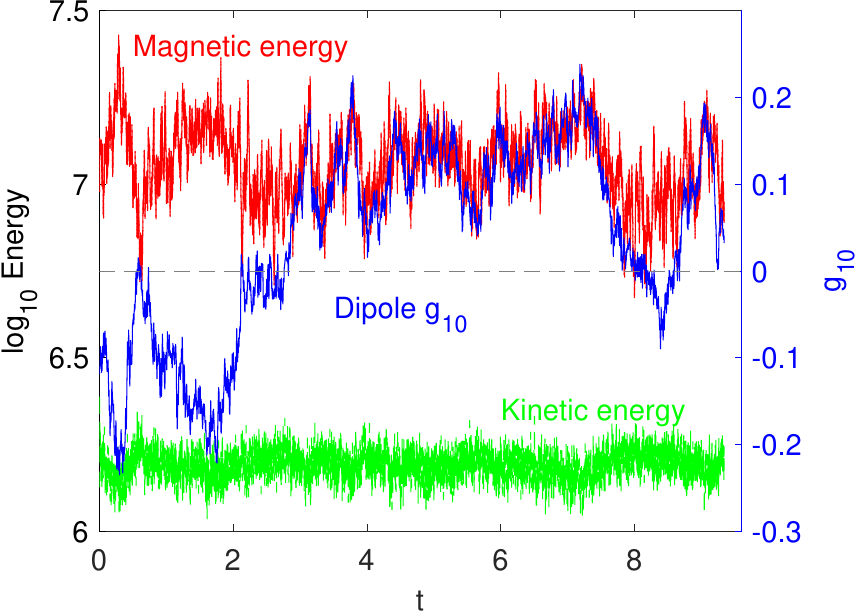} 
\end{minipage}

\caption{Magnetic energy, kinetic energy and axial dipole coefficient $g_{10}$ at the CMB as a function of magnetic
diffusion time for four Earth-like reversing dynamos. (a) run C3, $Ra=4.5 \times 10^7$, $Pr=35$, $Pm=40$, $E=2 \times 10^{-4}$, 
 (b) run D2, $Ra=3 \times 10^7$, $Pr=40$, $Pm=40$, $E=2 \times 10^{-4}$,  (c) run J8, $Ra=2.5 \times 10^8$, $Pr=60$, $Pm=40$, $E=10^{-4}$,  (d) run K2, $Ra=2 \times 10^8$, $Pr=70$, $Pm=40$, $E=10^{-4}$.}
\label{figs:fig4}
\end{figure*}

%
%
%
%

\begin{figure*}
\centering
\begin{minipage}{0.48\textwidth}
(a)\\
\includegraphics[width=0.95\textwidth]{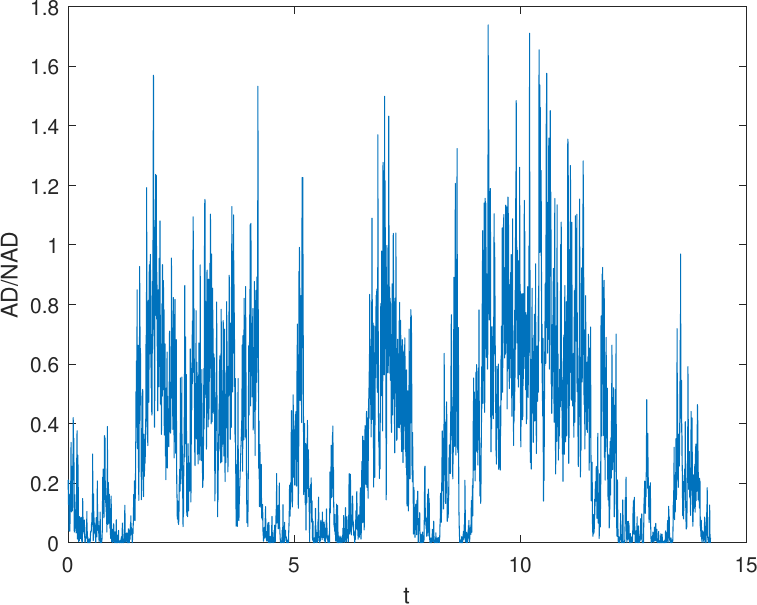}\\
(c)\\
\includegraphics[width=0.95\textwidth]{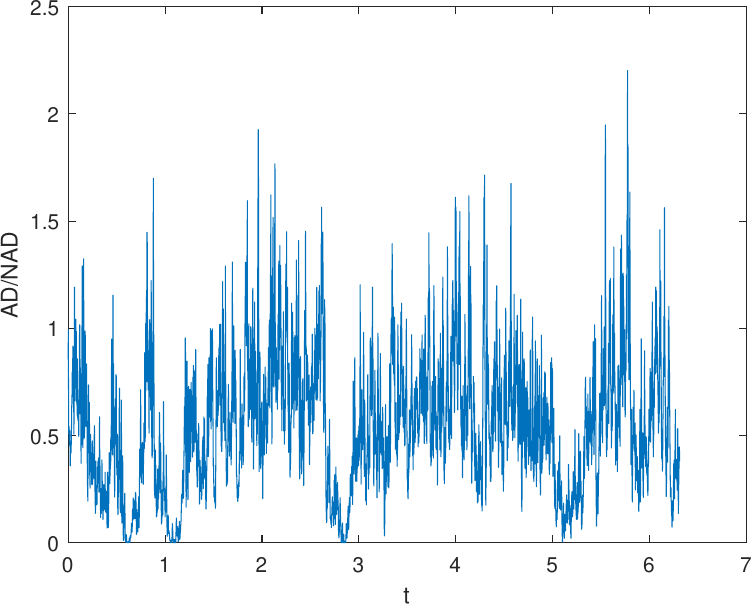}
\end{minipage}
\begin{minipage}{0.48\textwidth}
(b)\\
\includegraphics[width=0.95\textwidth]{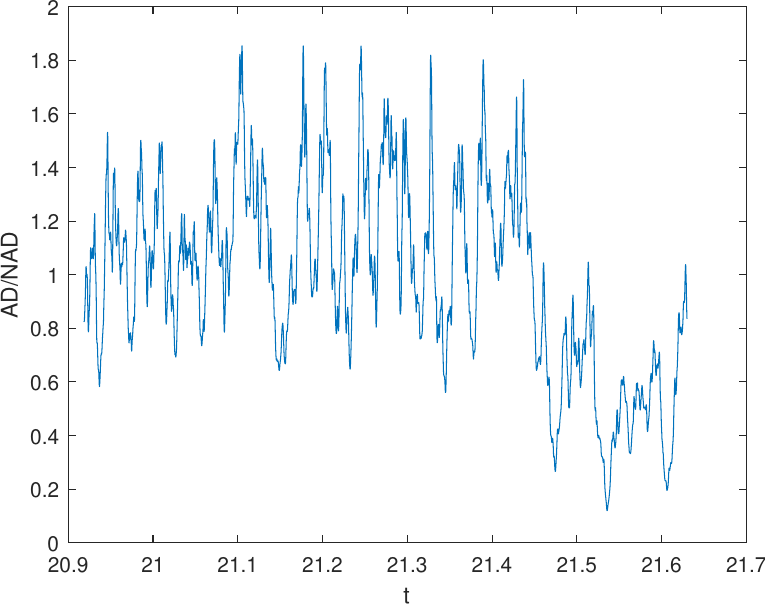}\\
(d)\\
\includegraphics[width=0.95\textwidth]{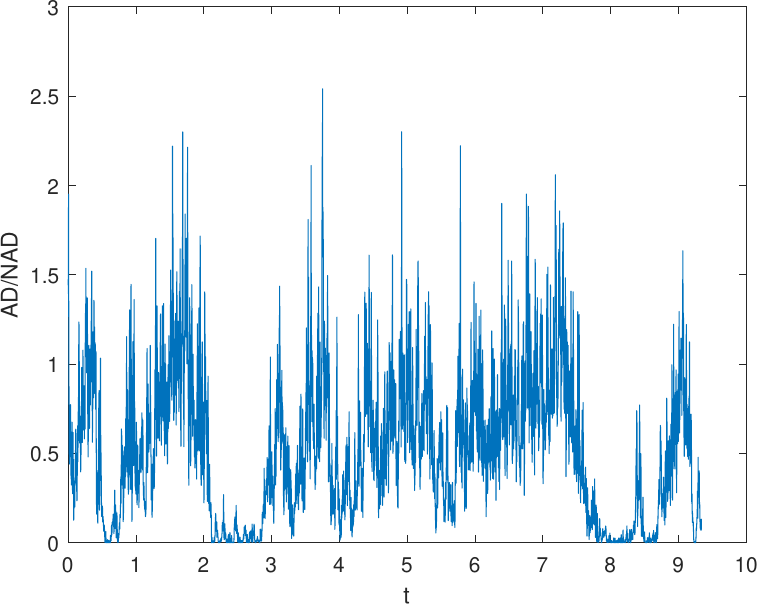} 
\end{minipage}

\caption{The relative axial dipole power, AD/NAD, as a function of magnetic
diffusion time for four Earth-like reversing dynamos. (a) run C3, $Ra=4.5 \times 10^7$, $Pr=35$, $Pm=40$, $E=2 \times 10^{-4}$, 
 (b) run D2, $Ra=3 \times 10^7$, $Pr=40$, $Pm=40$, $E=2 \times 10^{-4}$,  (c) run J8, $Ra=2.5 \times 10^8$, $Pr=60$, $Pm=40$, $E=10^{-4}$,  (d) run K2, $Ra=2 \times 10^8$, $Pr=70$, $Pm=40$, $E=10^{-4}$. }
\label{figs:fig5}
\end{figure*}

\subsection{Characteristics of the Earth-like reversing dynamos}

In Fig. \ref{figs:fig4} we show the magnetic energy (ME) and kinetic energy (KE) of some of the reversing Earth-like dynamos, on a logarithmic scale, with
the CMB axial dipole component $g_{10}$. The magnetic energy is larger than the kinetic energy in all these runs, as it is in the Earth, but the ratio ME/KE
is much smaller than in the real Earth. Both KE and ME have fluctuations on the turn-over time-scale, which is much shorter than the magnetic diffusion time-scale.
The fluctuations in the magnetic energy are relatively bigger than the fluctuations in the kinetic energy. Convective plumes can stretch out magnetic field to give locally very strong fields, even though the flow speeds in the plumes are not particularly big. The axial dipole component contributes a significant amount to the total magnetic energy,
and so times when $g_{10}$ is near zero correlate well with times of low ME.  

Fig. \ref{figs:fig5} plots the first \black{\cite{Christensen2010}} criterion for an Earth-like field, the axial dipole to non-axial dipole ratio, AD/NAD, see eq. (\ref{eq:ADNAD}). For the Earth's dynamo, this ratio is now around unity, but it was higher in the last few hundred years. Using archaeomagnetic and paleomagnetic data we find that the current dipole strength, and hence presumably AD/NAD, is larger than normal, so an Earth-like reversing dynamo should have periods when AD/NAD is greater than 1, but longer periods when it is less than one.
As we see in Fig. 5, most of our Earth-like dynamos satisfy this criterion. Note that only a chunk of the time-series for case D2 is shown, as this enables the fluctuations to be seen more clearly. All these dynamos have intervals where the AD/NAD ratio is similar to the current value, but the mean value is significantly less. 
The non-reversing dynamos can have significantly larger AD/NAD ratios, (compare the radii of the non-reversing dynamos in Fig. \ref{figs:fig3} with the reversing ones) but no reversing dynamos had a significantly larger AD/NAD ratio. Of the other criteria given in \black{\cite{Christensen2010}}, the ratio of the odd to even spherical harmonics O/E, which measures the asymmetry across the equator, and the ratio of the zonal to non-zonal coefficients, Z/NZ, give values similar to current Earth values for all the runs shown in Fig. \ref{figs:fig5}. As we see below, at full numerical resolution the radial magnetic field at the CMB is sometimes highly concentrated into intense flux patches, but when viewed at the resolution of only $l = 8$, defining the flux concentration factor criterion in \cite{Christensen2010}, these intense patches are no longer visible, so flux concentration is highly dependent on the resolution used.

\begin{figure*}
\hspace{10mm} (a) \hspace{70mm} (b)

\includegraphics[width=0.468\hsize]{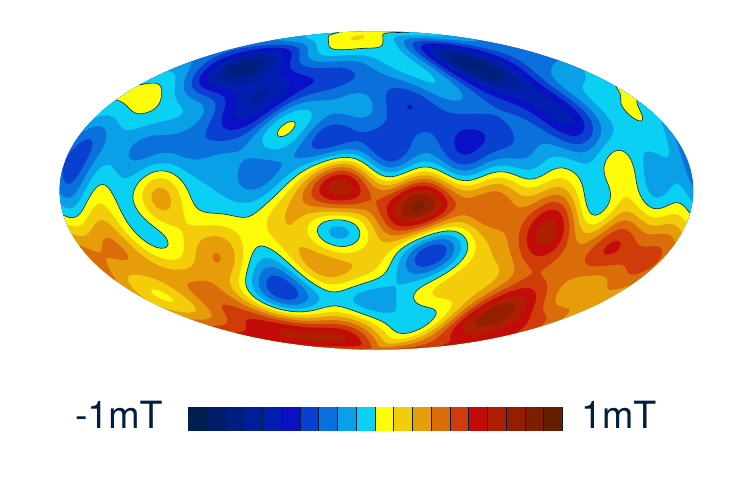} 
\hspace{1mm}
\raisebox{3mm}{\includegraphics[width=0.4\hsize]{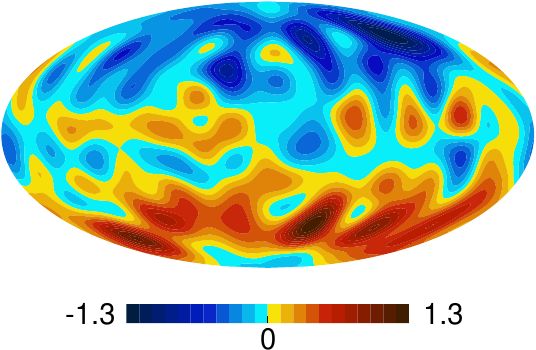}} 
\vspace{5mm}

\hspace{10mm} (c) \hspace{70mm} (d)

\hspace{2mm}\includegraphics[width=0.4\hsize]{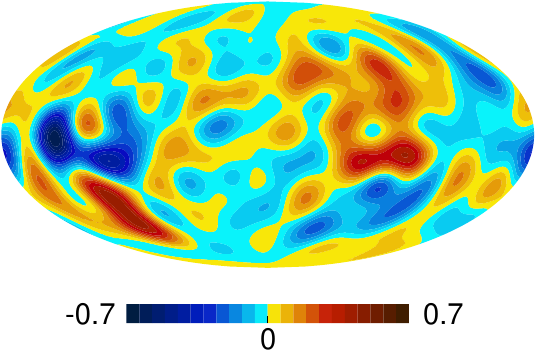} 
\hspace{10mm}
\includegraphics[width=0.4\hsize]{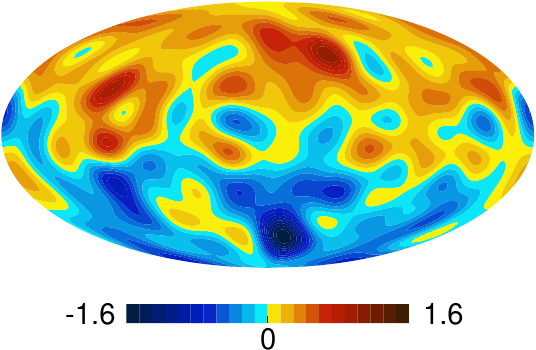} 
%
\caption{Snapshots of $B_r$ at the CMB at selected times for run D2, $Ra=3 \times 10^7$, $Pr=40$, $Pm=40$, $E=2 \times 10^{-4}$, compared with Earth's magnetic field in the year 2000. All plots have been truncated above $l = 13$ which are the degrees 
reliably known for the geomagnetic field. (a) The Geomagnetic $B_r$ at the CMB in the year 2000. (b) Run D2 at time 
$t=20.476$. (c) Run D2 at time $t=18.571$.  (d) Run D2 at time $t=16.588$.  }
\label{figs:fig6}
\end{figure*}

%


\subsection{Snapshots of the radial field at the CMB}

In Fig. {\ref{figs:fig6}} we compare the radial component of the magnetic field at the core-mantle boundary 
in the year 2000, Fig. {\ref{figs:fig6}}a, with the corresponding quantity at three different times from run D2, one of the models that had Earth-like reversing behaviour, Figs. {\ref{figs:fig6}}b,c,d. The units here
are dimensionless, but we can see that the unit of 0.85mT given in ({\ref{eq:Bunit}}) gives results compatible with the field strength in the year 2000. Fig. {\ref{figs:fig6}}b is taken at a time when the dipole is strongly negative, and the field, which has been truncated at degree 13, resembles the current field, broadly dipolar but with reversed field patches. Fig. {\ref{figs:fig6}c corresponds to a time when the field was reversing, and the dipole component is quite small. Apart from the lack of the dipole component the field is not so different from the strong dipole cases. It seems that there is no strong correlation between individual spherical harmonic components during the reversal process in these models, so a giant Gaussian process \citep{Constable1988} would appear to be a reasonable description of the behaviour of these models, though this has not been tested in detail.  Fig. {\ref{figs:fig6}}d is a snapshot of the field when the dipole was relatively large and positive, and as expected the field is broadly similar to that in  Fig. {\ref{figs:fig6}}b except that the sign of the field is reversed. 

In Fig. {\ref{figs:fig7}} we show the radial field of a lower $E$ reversing dynamo model J8 at two times with opposite polarity. Figs. {\ref{figs:fig7}}a,b are truncated above $l = 13$, and at the two times shown they have an AD/NAD ratio of 1.2 and 1.6 respectively. They also look reasonably Earth-like at these times. In Figs. {\ref{figs:fig7}}c,d the full resolution runs are shown at the same two times and they look remarkably different, with a great deal of small scale structure. Careful inspection shows that all the stronger field features in the high resolution pictures are present in the truncated snaphots, but they are smeared out by the truncation process so they look larger but weaker. Note that the peak field of the full resolution plots is three times the peak field of the truncated plots. The observed field has to be truncated at around degree $l = 13$ in order to eliminate the (unknown) contribution from remanent magnetism in the mantle, but these figures indicate that much fine structure in the field is lost during this process.

%
%
%
%

\begin{figure*}
\centering
\begin{minipage}{0.48\textwidth}
(a)\\
\includegraphics[width=0.95\textwidth]{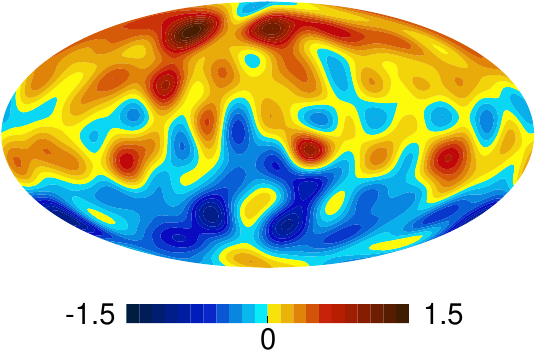}\\
(c)\\
\includegraphics[width=0.95\textwidth]{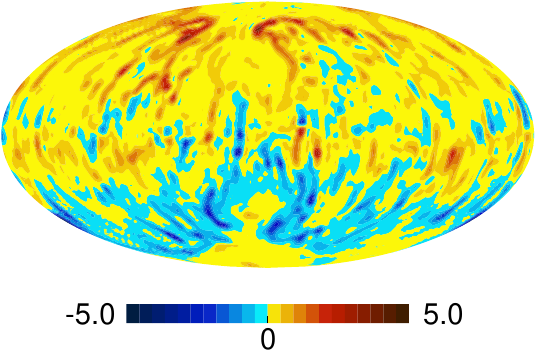} 
\end{minipage}
\begin{minipage}{0.48\textwidth}
(b)\\
\includegraphics[width=0.95\textwidth]{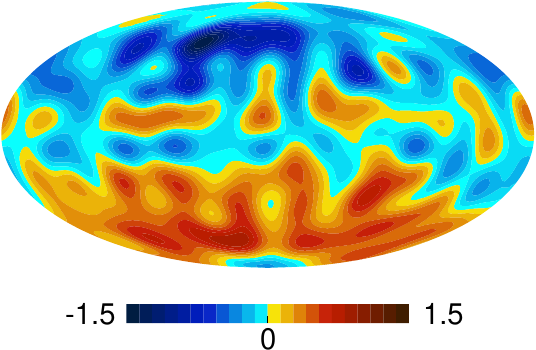}\\
(d)\\
\includegraphics[width=0.95\textwidth]{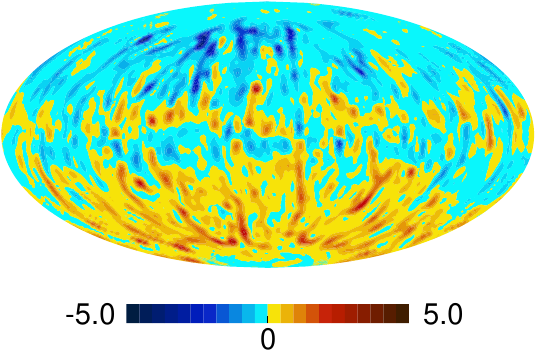} 
\end{minipage}

\caption{Snapshots of $B_r$ at the CMB at selected times for run J8, $Ra=2.5 \times 10^8$, $Pr=60$, $Pm=40$, $E=10^{-4}$. 
(a) Run J8 at time $t=2.1922$, truncated at $l=13$. (b) Run J8 at time 
$t=4.1364$ also truncated at $l=13$. (c) Same as (a) but with full resolution (no truncation).  (d) Same as (b) but  with full resolution.}
\label{figs:fig7}
\end{figure*}

%
%

\begin{figure*}
\centering
\begin{minipage}{0.48\textwidth}
(a)\\
\includegraphics[width=0.95\textwidth]{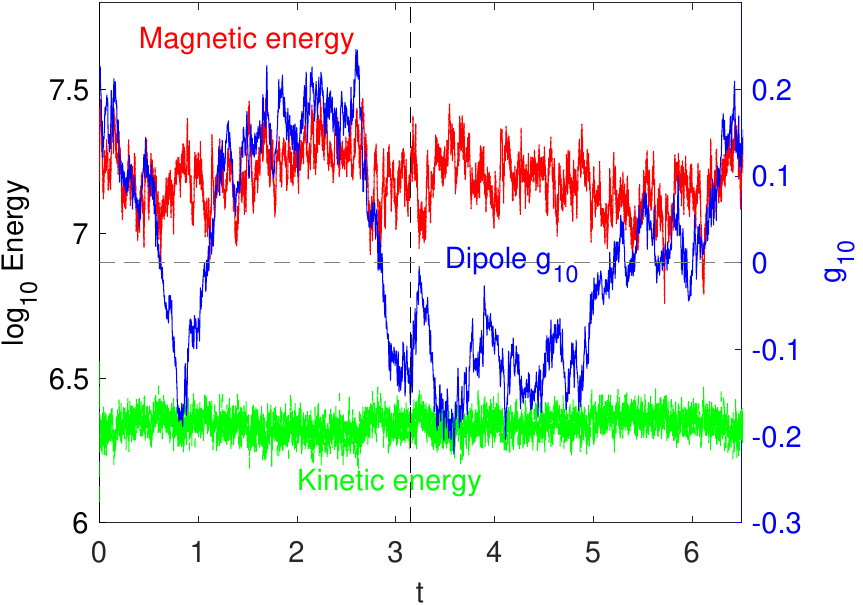} 
\end{minipage}
\begin{minipage}{0.48\textwidth}
(b)\\
\includegraphics[width=0.95\textwidth]{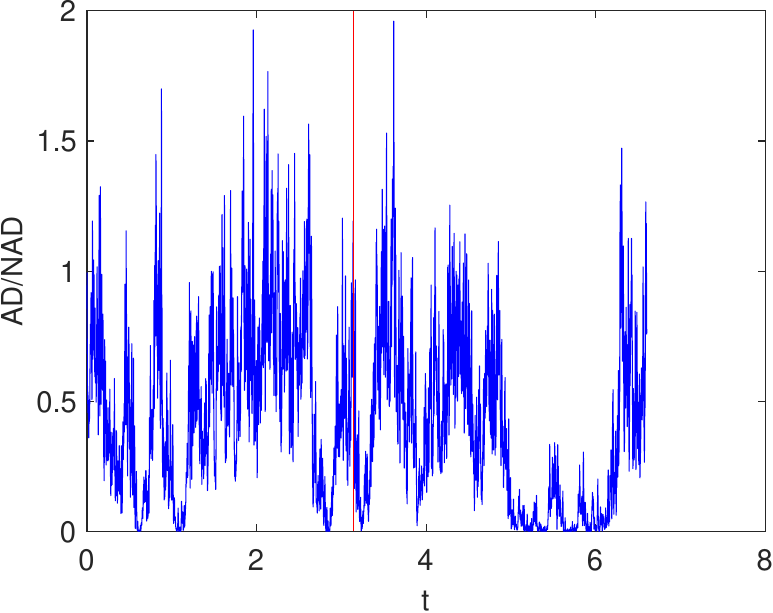} 
\end{minipage}

\caption{(a) Magnetic energy, kinetic energy and dipole $g_{10}$ at the CMB, and (b) the relative axial dipole power AD/NAD, both for the J8 case parameters. Up until $t=3.1461$ the resolution was $Nr=160$, $Nl=128$ and $Nm=128$. After $t=3.1461$, the run was restarted at a higher resolution with $Nl = 192$ \black{and $Nm=192$}. Despite the change in resolution, the behaviour of the system is not significantly different after $t=3.1461$.}
\label{figs:fig8}
\end{figure*}

%
%

\begin{figure*}
\centering
\begin{minipage}{0.48\textwidth}
(a)\\
\includegraphics[width=0.95\textwidth]{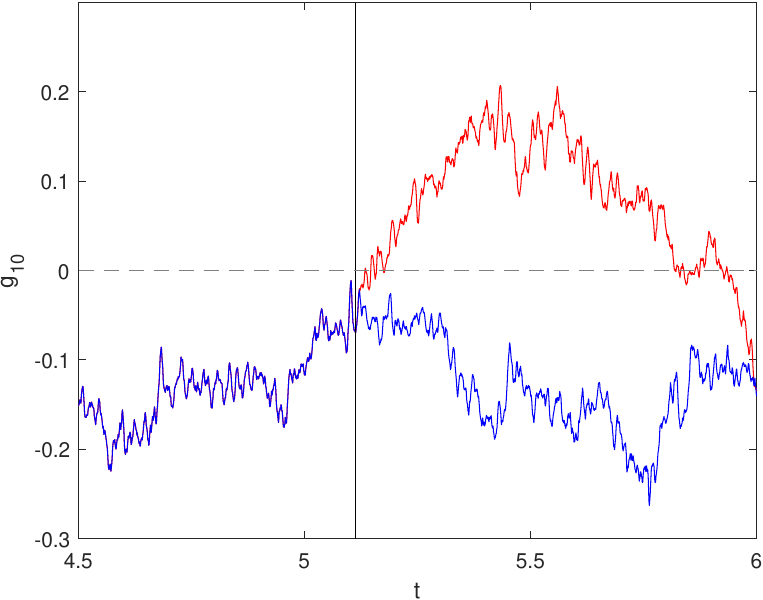} 
\end{minipage}
\begin{minipage}{0.48\textwidth}
(b)\\
\includegraphics[width=0.95\textwidth]{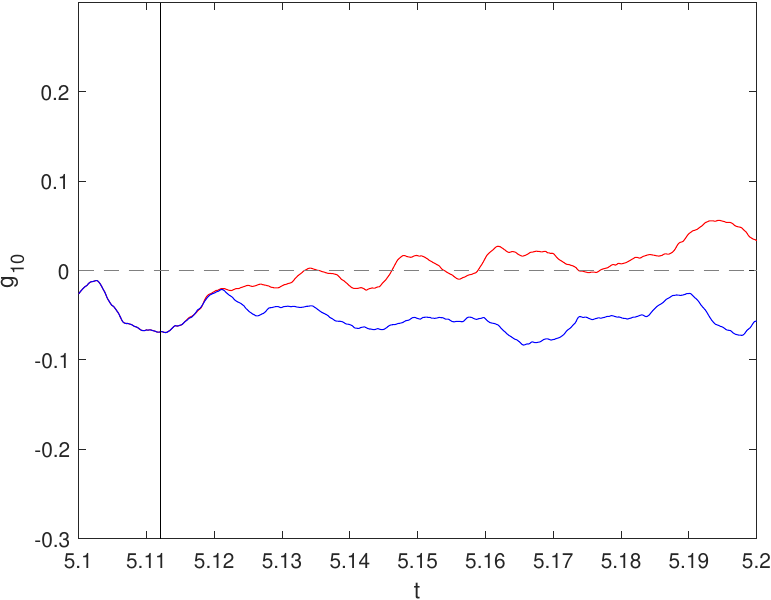} 
\end{minipage}

\caption{(a) The axial dipole $g_{10}$ at the CMB for the J8 case parameters. At $t=5.1121$ the run was interrupted and restarted, giving a very small perturbation to the system. After this time, the red curve follows the interrupted system with the perturbation, the blue curve follows the uninterrupted system. (b) is the same, but magnifies the timescale to see the transition region. Note that after  the restart, the two runs initially follow very similar trajectories, as the perturbation is very small, but they quickly diverge because the small perturbation grows exponentially corresponding to a positive Liapunov exponent. The tiny perturbation has led to a full reversal (red curve case) whereas the unperturbed blue curve case does not reverse then.}
\label{figs:fig9}
\end{figure*}

\begin{figure}
\centering
\includegraphics[width=1.0\hsize]{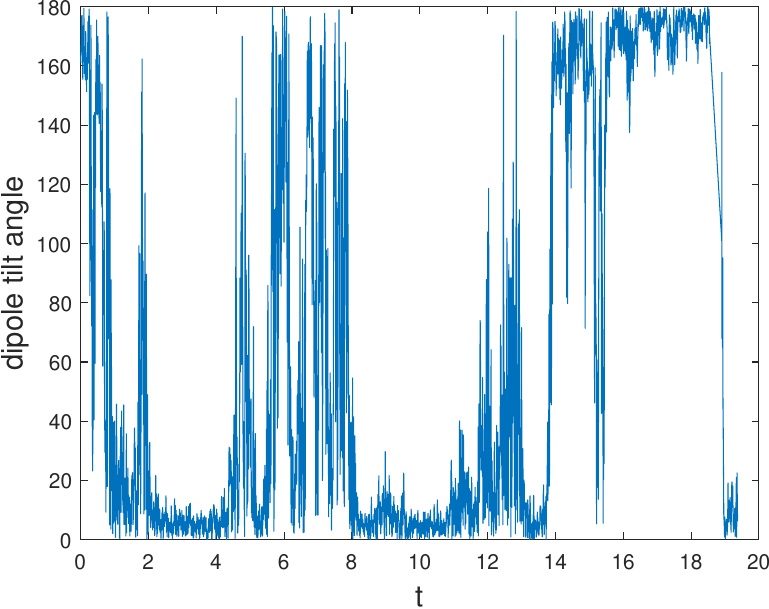} 
\caption{The axial dipole tilt angle from the MAGIC run. The MAGIC input parameters were $Ra=5 \times 10^7$, $Pr=Pm=40$, $E=2 \times 10^{-4}$. 
This is close to the case D2 values For the input parameters defined here. Note that there are reversals \black{in} this MAGIC run, which satisfy our criteria for an Earth-like reversal.}
\label{figs:fig10}
\end{figure}

\subsection{Numerical checks and randomness}

Because of the need for long runs to establish whether dynamos reverse or not, it is not possible to do every run at high resolution. 
We restricted our search of parameter space to those regions where the spectrum of the magnetic, velocity and thermal fields were all getting small at the highest wavenumbers computed. Hyperdiffusion was not used here, though it would be interesting to explore whether it could lead to computational savings without compromising the results. Most runs were done with a resolution of 160 radial nodes, and spherical harmonics up to degree $l=Nl$ and order $m=Nm$ both with $Nl=Nm=128$. The radial resolution was checked and looked satisfactory in all cases, but to check the spherical harmonic truncation, some runs were performed up to degree and order 192. The degree cut-off $Nl$ is the most critical, as the rate of convergence of the order $m$ is generally faster than that of the degree. Unfortunately, it is increasing the cut-off of the degree which is most burdensome computationally, even though the speed-up techniques developed by  \cite{Schaeffer2013} were used. 

In Fig. {\ref{figs:fig8}}, the case J8 was run up to time $t=3.1461$ with resolution $Nr=160$,
$Nl=128$ and $Nm=128$. At time $t=3.1461$ the run was continued with resolution $Nr=160$, $Nl=192$ and $Nm=192$. In Fig. {\ref{figs:fig8}}a we show the magnetic and kinetic energy and CMB dipole $g_{10}$, marking the transition time by a vertical dashed line. There is no noticeable change in the behaviour of any of these quantities after the transition time indicating that the original resolution was adequate for this case. In Fig. {\ref{figs:fig8}}b we see that similarly there is no noticeable change in the behaviour of the ratio AD/NAD with increased resolution. 

Despite the fact that our system is deterministic, with no random forcing in the equations, trajectories which are initially close together rapidly diverge, so the chaotic nature of the system means that there is considerable randomness in practice. To illustrate this
behaviour, we performed a run with the case J8 parameters up to time $t=5.1121$. We then restarted the  run twice at $t=5.1121$. On the first rerun, we used exactly the same starting solution as we ended up with from the run which finished at $t=5.1121$ and the blue curves in Figs. {\ref{figs:fig9}}a,b follow these trajectories. For the second restarted run, a small perturbation of order $10^{-6}$ was added to the magnetic field variables. The red curves show the dipole $g_{10}$ for this second restarted run. In Fig. {\ref{figs:fig9}}b
we see the two different runs track each other initially after $t=5.1121$, as expected because the perturbation is so small. However,
the two curves soon start to diverge, and they are noticeably different after $t=5.12$. This is perhaps less surprising when we note that the convective turn-over time, $1/Rm$, is only $t \approx 0.002$ in these units, so divergence has become noticeable after 4 eddy turn-over times. However, the two trajectories continue to diverge, so looking at \black{Fig.} {\ref{figs:fig9}}a we see that after the perturbation the red curve has reversed the sign of $g_{10}$ while the original blue curve did not. So this tiny perturbation has led to a full reversal, whereas the unperturbed solution did not reverse. Since all computers have tiny rounding errors due to the finite number of digits stored, this means that small perturbations are unavoidable and randomness is built in, even though our system is in theory completely deterministic.

We also solved our equations using the MAGIC package (https://github.com/magic-sph/magic) to check that the reversal behaviour was similar to that obtained by the Leeds code. We looked at the case D2 in table 1, which has Earth-like reversals. The definition of the Rayleigh number has a factor $(r_0-r_i)/r_0$ different from eq. ({\ref{eq:Ra_defs}}), so we kept $E=2 \times 10^{-4}$, $Pr=Pm=40$ but changed the
MAGIC input Rayleigh number to $5 \times 10^7$ to compensate for this factor. The behaviour obtained from this MAGIC run had similar values of magnetic  and kinetic energy and magnetic Reynolds number to those of the Leeds code runs. In Fig. 10 we show the dipole tilt angle for a run of 20 magnetic diffusion times, a similar length to run D2. It is clear that the MAGIC run gives reversals very similar to those found in the $g_{10}$ plot in Fig. \ref{figs:fig4}b. 


%
%
%
%

\begin{figure*}
\centering
\begin{minipage}{0.48\textwidth}
(a)\\
\includegraphics[width=0.95\textwidth]{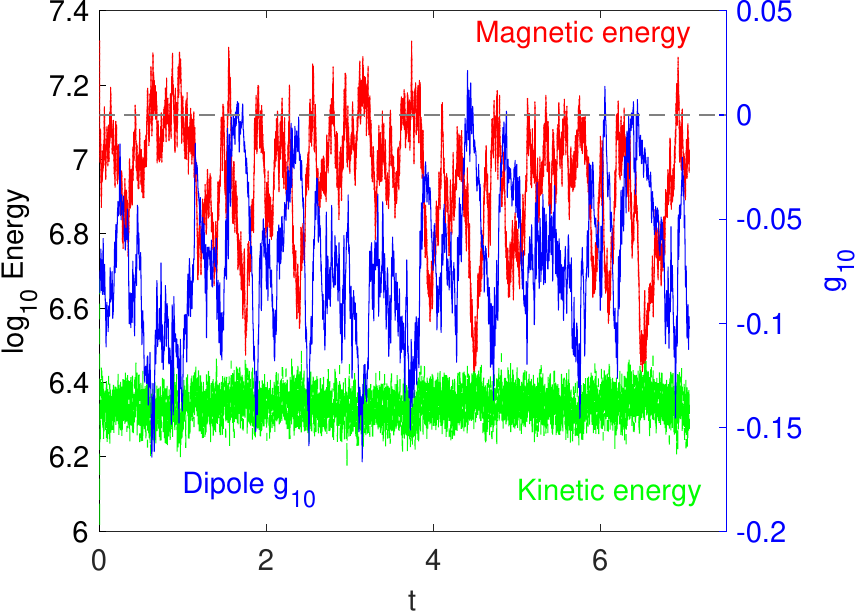}\\
(c)\\
\includegraphics[width=0.95\textwidth]{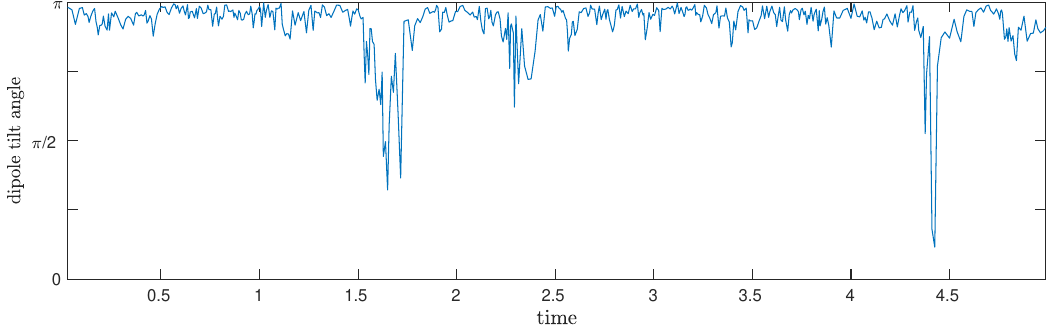}
\end{minipage}
\begin{minipage}{0.48\textwidth}
(b)\\
\includegraphics[width=0.95\textwidth]{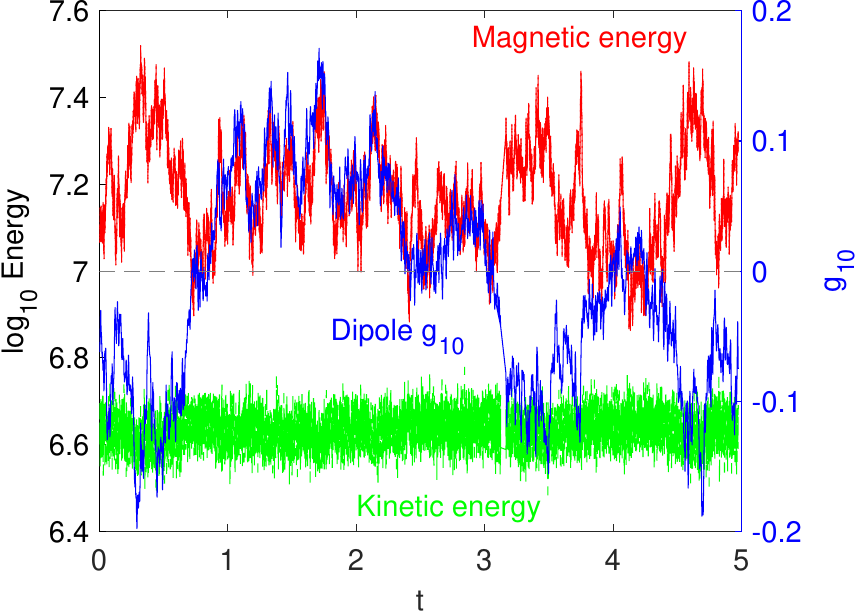}\\
(d)\\
\includegraphics[width=0.95\textwidth]{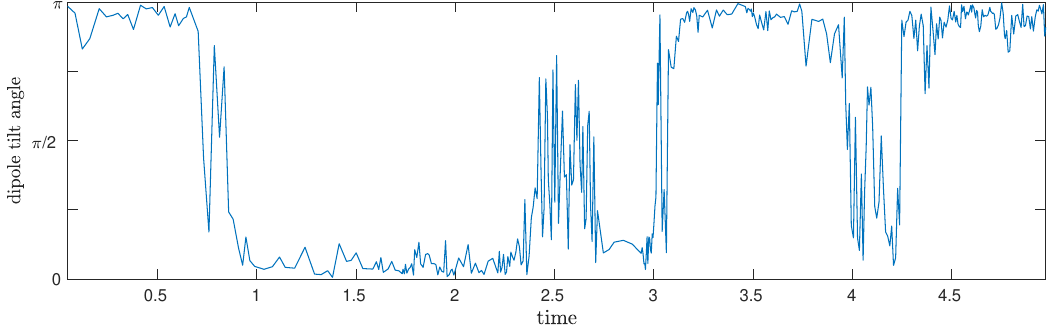} 
\end{minipage}

\caption{Compositional convection runs. (a) Magnetic energy, kinetic energy and dipole coefficient $g_{10}$ at the CMB for run \black{CCB3} with $Pr=40$, $Pm=20$, $E=7 \times 10^{-5}$, $Rm=545$: (c) Dipole  axis angle for the same run. This run has excursions, but no full reversals.
(b) Magnetic energy, kinetic energy and dipole coefficient $g_{10}$ for run \black{CCB1} with $Pr=40$, $Pm=20$, $E=7 \times 10^{-5}$, $Rm=766$: (d) 
Dipole  axis angle for the same run. This run, at higher $Ra$ and $Rm$, has full reversals. }
\label{figs:fig11}
\end{figure*}

%
%
%
%
%
%

\begin{figure*}
\centering
\begin{minipage}{0.48\textwidth}
(a)\\
\includegraphics[width=0.90\textwidth]{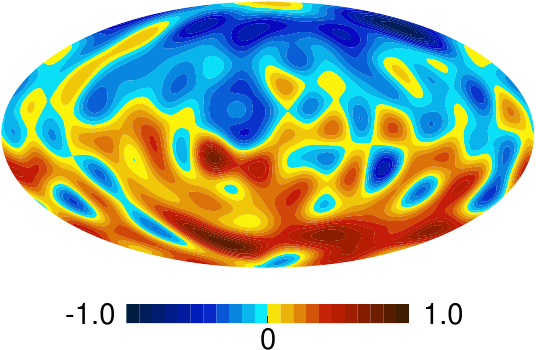}\\
(c)\\
\includegraphics[width=0.9\textwidth]{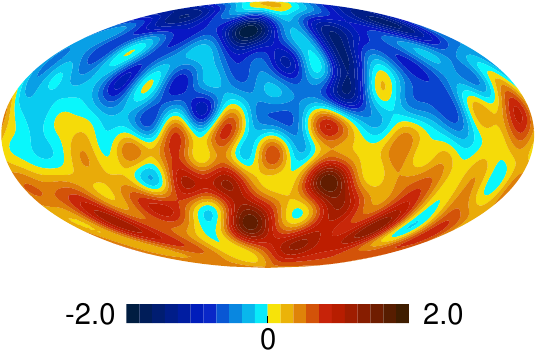}\\
(e)\\
\includegraphics[width=0.95\textwidth]{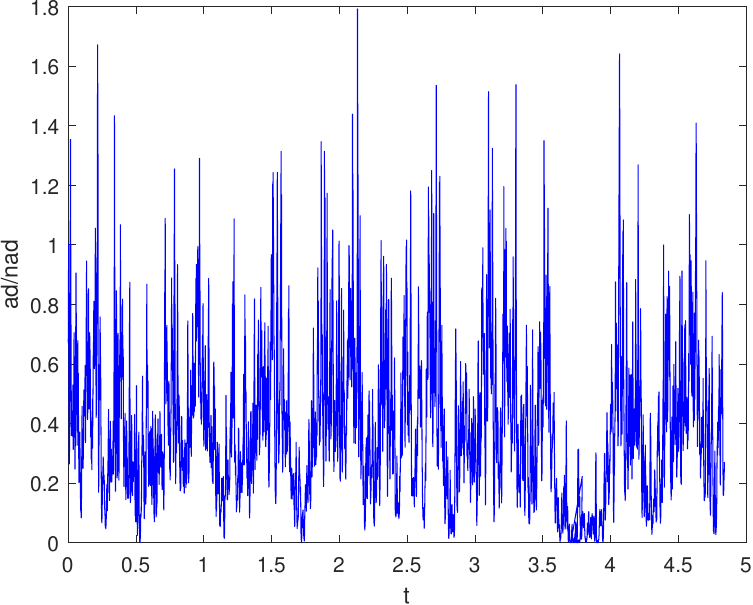} 
\end{minipage}
\begin{minipage}{0.48\textwidth}
(b)\\
\includegraphics[width=0.90\textwidth]{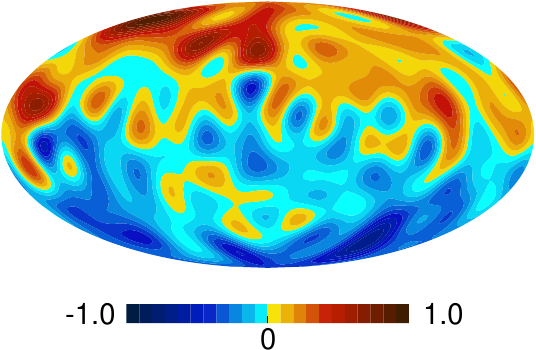}\\
(d)\\
\includegraphics[width=0.9\textwidth]{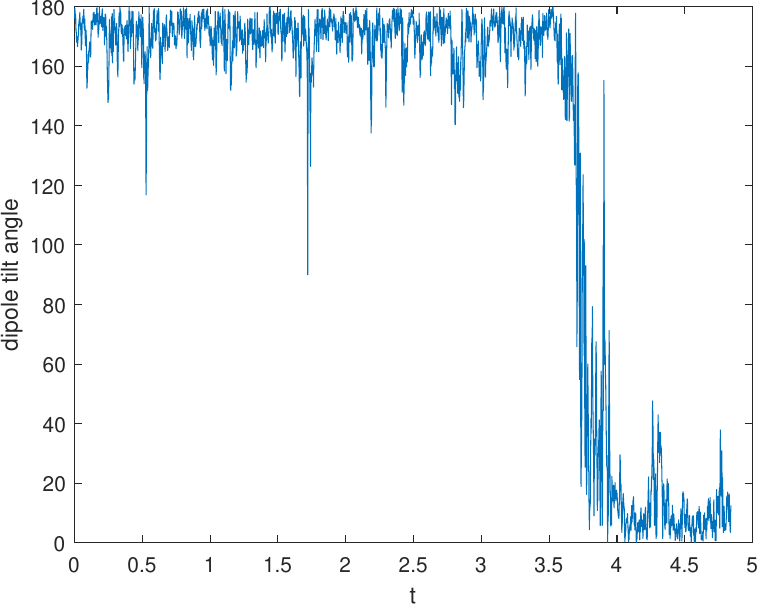}\\
(f)\\
\includegraphics[width=0.90\textwidth]{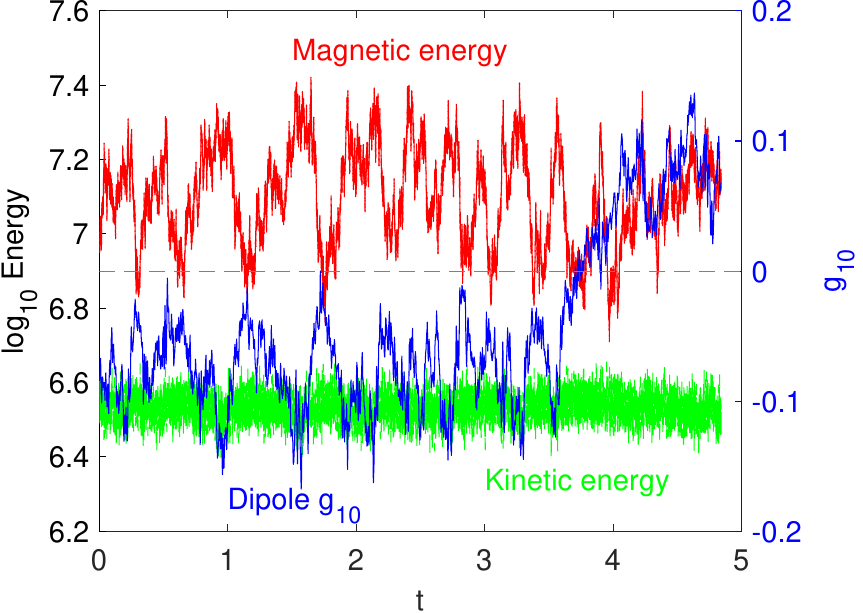} 
\end{minipage}

\caption{Compositional convection runs. (a,b) Snapshots of $B_r$ at the CMB at selected times for run CCB4, $Ra=2 \times 10^9$, $Pr=35$, $Pm=20$, $E=7 \times 10^{-5}$. (a) $t=2.134$, (b) $t=4.631$ after the reversal. (c) Snapshot of $B_r$ at  the CMB for run CBB0, $Ra=6.75 \times 10^{8}$, $E=2.5 \times 10^{-5}$, $Pr=1$, $Pm=2.5$.   All the fields in figures (a-c) have been truncated at $l=13$. (d) Dipole tilt angle for run CCB4. (e) Axial dipole to non-axial field ratio, AD/NAD for run CCB4. (f) Magnetic and kinetic energies for run CCB4 together with the axial dipole component $g_{10}$ at the CMB.}
\label{figs:fig12}
\end{figure*}

\section{Results from the dynamos driven by compositional convection}

Figs.~\ref{figs:fig11}a and \ref{figs:fig11}b} show the timeseries of kinetic and magnetic energy and the dipole strength $g_{10}$ of the two simulations CCB3 and CCB1 (see table 3) using the CCB boundary conditions (\ref{eq:CCB}). Earth-like reversals were harder to find with the CCB boundary conditions than with the fixed flux FFFF boundaries. These two runs had a smaller $E$, only $7 \times 10^{-5}$, than the FFFF runs. Even with a larger $Rm=545$ and $Pe=1090$, run CCB3 \black{did not} have any Earth-like reversals. In Fig. \ref{figs:fig11}c the dipole tilt for CCB3 is shown. Although there are several excursions (where the field nearly reverses but \black{does not} quite manage it) there are no true Earth-like reversals in the series. However, when $Ra$ and hence $Rm$ is increased so that $Rm=766$ and $Pe=1532$, run CCB1, Earth-like reversals do occur, see Fig. \ref{figs:fig11}d. Both these runs have a significantly larger magnetic energy than kinetic energy, so that the effects of inertia are relatively weak. Unfortunately the larger $Rm$ and smaller $E$ mean that a higher truncation is necessary,  mostly  \black{160 $\times$ 192 $\times$ 192}, though as the $Nm$ truncation is not quite so critical as the $Nl$ truncation, run CCB4 used \black{160 $\times$ 192 $\times$ 128,} thus allowing a longer run than would otherwise be possible. 

We see two reversals in Run CCB1. The first one near the beginning at $t\approx0.8$ may or may not be a transient effect. Somewhere between $t=2$ and $t=2.5$, the dynamo starts transitioning from a positive dipolar state to a  \black{multipolar} state, and then at around $t=3$, it reverses into a negative dipolar state. The length of the transition period is in general fairly short at about $0.2$ while the dipolar and the \black{multipolar} states are sustained for an interval of about 0.5 or longer. If the reversal at $t=0.8$ is indeed a transient effect, then it is not clear whether the dynamo can reverse its polarity without first going through a  \black{multipolar} state. In addition to the reversals at $t=0.8$ and $t=3$, the dynamo also undergoes an excursion at about $t=4$. During this excursion, the negative dipolar state collapses into a \black{multipolar} state at $t=3.8$ before re-emerging at $t=4.3$.

Run CCB2 is not shown,  because it was  \black{multipolar} all the time. With these boundary conditions, we explored the parameter space between $E=1 \sim 2 \times 10^{-4}$ but found no convincing Earth-like reversing dynamos, different from the FFFF case where several were found. It seems that lower $E$ and higher $Rm$ is needed for Earth-like reversals with compositional convection boundary conditions. It is possible, but very hard to test, that Earth-like reversing dynamos are just as plentiful with CCB conditions, but they exist in a more challenging region of the parameter space. 

Run CCB1 did exhibit Earth-like reversals, but its AD/NAD ratio rarely got above 1, so that although it reverses in a satisfactory manner, the field in the dipole state is only very rarely Earth-like. From our experience with the FFFF cases, run CCB4 with a small reduction in Pr from 40 to 35 was tried, slightly increasing $q$ and decreasing $Pe$. This did indeed increase the AD/NAD ratio a little, and the results are displayed in Fig. 12. As can be seen from Fig. \ref{figs:fig12}d this dynamo did reverse  reasonably cleanly around $t \sim 3.6$. Only one reversal was found, which is not surprising as $Rm = 683$ and $Pe=1195$ so the fluctuation level is a little weaker than in run CCB1, which reversed more frequently.

These compositional convection dynamos are similar to the FFFF dynamos, in that the magnetic energy is consistently larger than the kinetic energy (see Fig. \ref{figs:fig12}f) and the fluctuations in the magnetic field are much larger than the fluctuations in the kinetic energy. In Fig. \ref{figs:fig12}e the AD/NAD ratio of run CCB4 is shown. There are now significant intervals when AD/NAD exceeds unity. Snapshots taken in these intervals are shown in Figs. \ref{figs:fig12}a,b showing that field has indeed reversed, and the states achieved in these intervals show reasonably Earth-like fields. For comparison, we looked at a case with conventional Earth-like field parameters, run CCB0 with $Pr=1$ and $Pm=2.5$ as chosen in \cite{Christensen2010}. The results of this simulation are shown in Fig. \ref{figs:fig12}c. It is significantly more dipolar than the fields shown in \ref{figs:fig12}a,b. Although $Rm$ is only a little smaller, $Pe$ is much smaller. Run CCB0 showed no sign of reversing. Increasing $q$ and therefore reducing $Pe$ seems to have the effect of increasing the AD/NAD ratio, but it also decreases the fluctuation levels and hence makes the dynamo reverse less often. Reversals are rare in the current state of the geodynamo, so it is possible more realistic behaviour could be achieved by tweaking the parameters, but unfortunately longer intervals between reversals greatly adds to the computational expense. 


\section{Discussion and Conclusions}

To be consistent with the paleomagnetic data, and hence considered Earth-like, the dynamo should not only reverse polarity, i.e. the sign of the axial dipole $g_{10}$ must change, there should also be long periods when the dynamo is dipole dominated, defined here as an AD/NAD ratio greater than unity. In view of the existence of hyper-active periods of magnetic activity \citep{Gallet2019}, windows of multipolar behaviour might be considered acceptable, but not dynamos which reverse all the time. By looking at models where the fluid Prandtl number as well as the magnetic Prandtl number is relatively large, we have found 12 new reversing dynamos which satisfy these criteria and also have the feature that the magnetic energy is much larger than the kinetic energy and the Rossby number is small. The dynamos were found in the range of $7 \times 10^{-5} \le E \le 2 \times 10^{-4}$
for both types of boundary conditions considered, and with Roberts number $0.5 \le q \le 1.143$. 
It is therefore possible to have dynamos where the reversal mechanism is not greatly influenced by inertia, supporting the ideas of \cite{Sreenivasan2014}. This gives further support to the idea that the Earth has a convection driven dynamo. One of the objections to this hypothesis, that dynamo models only reverse at moderate Rossby number whereas the real Earth has very low Rossby number and yet reverses, has been overcome. 

\black{Our new reversing dynamos all have a much larger Ekman number than the Earth's core, from computational necessity, but the balance of terms in these simulations does reflect what is believed to be the correct balance in the core. It is therefore possible that despite the small scale modes being overdamped by excess diffusion, the modes driving the dynamo that we retain in our simulations may reflect what is driving the Earth's magnetic field. This may be why it is possible to find models that satisfy both the \cite{Christensen2010} criteria for an Earth-like magnetic field and produce Earth-like reversal behaviour.}

\black{Our models fall into three groups, the fixed flux models at $E=2 \times 10^{-4}$, the fixed flux models at $E=10^{-4}$ and the compositional convection models. The $E=2 \times 10^{-4}$ Earth-like reversing models have the advantage that they are less computationally costly, so they can be run for many diffusion times and their properties explored more fully. They do have the disadvantage that they are only found in a limited region of parameter space. If $q$ is too large, the dynamos do not reverse, if $q$ is too small the dynamos are multipolar. The $E=10^{-4}$ Earth-like FFFF dynamos are more expensive to run, but they have the advantage that they exist in a wider range of parameter space, so models which fit the past behaviour of the geomagnetic data can be selected. Like the $E=2 \times 10^{-4}$ dynamos there is a cut-off at larger $q$ beyond which the dynamos do not reverse (at least until large $Ra$ and $Rm$ are reached). At the lower boundary of $q$ for the $E=10^{-4}$ Earth-like reversing models, the dynamos stopped reversing, unlike the $E=2 \times 10^{-4}$ models which became multipolar. It is not yet clear how this difference arises. It is likely that compositional convection provides the main driving for the geodynamo at present, but unfortunately the Earth-like reversal regime was only found at $E=7 \times 10^{-5}$ in
the CCB models. This made it difficult to explore the CCB case in detail.}

The traditional criterion for the existence of Earth-like reversing dynamos, that the local Rossby number $Ro_l \sim 0.1$, appears to play no role in determining whether our $q \sim O(1)$ dynamos
are Earth-like or not. It seems that this local Rossby number criterion is specific to the choice
of Prandtl number $Pr=1$. 

\black{Since our models are chaotic, with trajectories that are very sensitive to initial conditions, it is natural to compare them with models which include explicit stochastic terms. The reversal behaviour in our models is dependent on the level of fluctuations, those with bigger fluctuations being more likely to reverse, so in this respect our models resemble that of \cite{Schmitt2001} and its successors. However, stochastic models usually have a simple imposed random forcing. In our model the fluctuations
have a spectrum in time and space, and it may be the fluctuation level at particular wavelengths in space and time that leads to reversals, so that models with the largest overall fluctuations are not necessarily those which reverse most often. }

\black{Another feature of the stochastic models is that they assume that the potential that controls the \lq particle' has a quartic form  with symmetric minima (see e.g. figure 2 of \cite{Schmitt2001}). While it is true that our models that have a low fluctuation level sit in an approximately dipolar state which can be of either sign, there is evidence that other types of attractor may exist. In some models, the trajectories spend time in small amplitude fluctuations about the $g_{10}=0$ state, multipolar behaviour, e.g run E4 between t=7 and t=9, run CCB1 between t=2.3 and t=2.7. This could be modelled in
the stochastic framework by having a potential with a local minimum at $g_{10}=0$ rather than the local maximum of \cite{Schmitt2001}. However, this does raise the question of how the potential in a stochastic model is to be chosen when the paleomagnetic record lacks comprehensive coverage. }

The way in which our low inertia reversing dynamos behave bears strong similarities to the behaviour at $Pr=1$, in that in order to provoke a reversal $Rm$ has to be increased so that the fluctuations in the field are large enough to overcome the barrier which keeps the field dipole dominated. However, in these new models it appears that it is not just the magnetic Reynolds number that plays a role, but the P\'eclet number also has an influence. When the nonlinear term in the codensity equation becomes large, the convection becomes more chaotic, thus  enhancing the fluctuations. It is still the case that the fluctuations in the magnetic field are larger than the fluctuations in the convection, because rising and falling plumes stretch out loops of magnetic field giving local large bursts of magnetic energy. The compositionally driven models of \cite{Schaeffer2017} showed particularly strong fluctuating activity inside the tangent cylinder. We believe that this is the case for our Earth-like reversing dynamos too, but
more work is needed to establish this conclusively. The FFFF dynamos, where reversals occurred at lower $Rm$ and $Pe$, are more strongly driven near the CMB whereas the CCB models are driven mainly near the ICB, which might explain why the CCB models had more excursions than reversals compared to the FFFF models. This could be consistent with meridional circulation advecting field across the equator more easily in the FFFF case, following the ideas of \cite{Wicht2004} who emphasised the importance of meridional circulation in promoting reversals.


We have not yet analysed the force balances in our simulations, but doing so would be of interest. As noted by \cite{Schwaiger2021} and \cite{Teed2023}, the force balance depends on the length-scale under consideration. On the length-scale of the core, a few thousand km, inertia is small compared to the Coriolis force, but at the Rhines length-scale, a few
km, they are comparable, and at even smaller scales the fluid hardly feels the effect of rotation. At these length scales there is likely to be a turbulent cascade down to the Kolmogorov scale at which the viscous dissipation occurs. It is possible that in this range there is an effective eddy diffusion, much larger than the molecular diffusion, and comparable to  the magnetic diffusion. On this picture, the effective value of $q$ could be of order unity rather than the very small values indicated by the molecular diffusion coefficients. In this situation, one might envisage a family of solutions, starting with the moderate $E$, high $Pr$ and $Pm$ solutions discussed here, in which $E$, $Pm$ and $Pr$ gradually reduce,
keeping $q \sim O(1)$, producing a larger and larger range of length-scales in the flow, until eventually the smallest resolved length-scales
would be at the Rhines scale, beyond which magnetic diffusion acts so fast that even smaller length scales would be irrelevant to the dynamo. The Rayleigh number in this family would be chosen so that the magnetic Reynolds number remained fairly constant, perhaps increasing weakly to the most plausible Earth-like value of O(1000). There is no possibility of establishing this with the numerical models used here, but it may be possible to develop reduced asymptotic models which speed up the computational process to allow  some progress to be made towards smaller $E$ solutions.


\begin{paragraph}
{\noindent {\bf Acknowledgements} \newline \rm
The authors were supported by the Science and Technology Facilities Council (STFC), `A Consolidated Grant in Astrophysical Fluids' (grant numbers ST/K000853/1 
and ST/S00047X/1). This work used the DiRAC Complexity system, operated by the University of Leicester IT Services, which forms part of the STFC DiRAC HPC Facility (www.dirac.ac.uk). This equipment is funded by BIS National E-Infrastructure capital grant ST/K000373/1 and  STFC DiRAC Operations grant ST/K0003259/1. DiRAC is part of the National E-Infrastructure.}
The authors would like to thank the Isaac Newton Institute for Mathematical Sciences, Cambridge, for support and hospitality during the programme DYT2 where part of this paper was written. DYT2 was supported by EPSRC grant EP/R014604/1.

\end{paragraph}

\vspace{2mm}
{\noindent \bf Data availability}\\
The numerical code used for the geodynamo simulations reported in this paper is at https://github.com/Leeds-Spherical-Dynamo. For access to the Github repository, please contact the authors. The MAGIC is available at
https://github.com/magic-sph/magic and is open access, with code documentation available at the github.com site.

\begin{table*}
 \begin{center}
\def~{\hphantom{0}}
\begin{tabular}{ccccccccccc}
 Run  & $Pr$ & $Ra$ & $\cal{R}$ & $q$ & $Rm$ & $Pe$ & $Dip$  & $Ro$ & $\tau$ & Reversal \\[3pt]
 
\hline
 A1 & 20 & $2.5 \times 10^7$ & 10000 & 2 & 354 & 177 & 0.105 & $1.77 \times 10^{-3}$ & 3.03 & N \\
 A2 & 20 & $4.0 \times 10^7$ & 16000 & 2 & 450 & 225 & 0.087 & $2.25 \times 10^{-3}$ & 2.24 & N \\
 B1 & 30 & $3.0 \times 10^7$ & 8000 & 1.333 & 331 & 248 & 0.101 & $1.66 \times 10^{-3}$ & 5.86 & N \\
 B2 & 30 & $5.0 \times 10^7$ & 13333 & 1.333 & 440 & 330 & 0.060 & $2.20 \times 10^{-3}$ & 8.35 & N* \\
 C1 & 35 & $3.0 \times 10^7$ & 6857 & 1.143 & 315 & 276 & 0.095 & $1.57 \times 10^{-3}$ & 8.97 & N \\
 C2 & 35 & $4.0 \times 10^7$ & 9143 & 1.143 & 364 & 318 & 0.157 & $1.82 \times 10^{-3}$ & 3.35 & N \\
 C3 & 35 & $4.5 \times 10^7$ & 10286 & 1.143 & 404 & 353 & 0.051 & $2.02 \times 10^{-3}$ & 14.23 & E \\
 C4 & 35 & $5.0 \times 10^7$ & 11429 & 1.143 & 424 & 371 & 0.049 & $2.12 \times 10^{-3}$ & 11.17 & M \\
 C5 & 35 & $6.0 \times 10^7$ & 13714 & 1.143 & 465 & 407 & 0.035 & $2.32 \times 10^{-3}$ & 5.87 & M \\
 D1 & 40 & $2.0 \times 10^7$ & 4000 & 1 & 239 & 239 & 0.119 & $1.19 \times 10^{-3}$ & 10.89 & N \\
 D2 & 40 & $3.0 \times 10^7$ & 6000 & 1 & 315 & 315 & 0.073 & $1.58 \times 10^{-3}$ & 20.94 & E \\
 D3 & 40 & $4.0 \times 10^7$ & 8000 & 1 & 361 & 361 &  0.063 & $1.80 \times 10^{-3}$ & 11.09 & E \\
 D4 & 40 & $4.5 \times 10^7$ & 9000 & 1 & 394 & 394 & 0.033 & $1.97 \times 10^{-3}$ & 2.46 & M \\
 D5 & 40 & $5.0 \times 10^7$ & 10000 & 1 & 401 & 401 & 0.041 & $2.01 \times 10^{-3}$ & 10.71 & M \\
 D6 & 40 & $6.0 \times 10^7$ & 12000 & 1 & 437 & 437 & 0.030 & $ \times 10^{-3}$ & 7.71 & M \\
 D7 & 40 & $8.0 \times 10^7$ & 16000 & 1 & 484 & 484 & 0.037 & $2.41 \times 10^{-3}$ & 3.60 & M \\
 E1 & 50 & $1.0 \times 10^7$ & 1600 & 0.8 & 155 & 194 &  0.151 & $0.77 \times 10^{-3}$ & 9.17 & N \\
 E2 & 50 & $1.2 \times 10^7$ & 1920 & 0.8 & 168 & 210 & 0.145 & $0.84 \times 10^{-3}$ & 4.38 & N \\
 E3 & 50 & $1.25 \times 10^7$ & 2000 & 0.8 & 200 & 250 & 0.047 & $1.00 \times 10^{-3}$ & 11.09 & M \\
 E4 & 50 & $1.5 \times 10^7$ & 2400 & 0.8 & 217 & 271 & 0.055 & $1.09 \times 10^{-3}$ & 9.28 & M \\
 E5 & 50 & $2.0 \times 10^7$ & 3200 & 0.8 & 247 & 309 & 0.052 & $1.23 \times 10^{-3}$ & 7.34 & M \\
 E6 & 50 & $3.0 \times 10^7$ & 4800 & 0.8 & 299 & 374 & 0.027 & $1.49 \times 10^{-3}$ & 5.62 & M \\
 E7 & 50 & $4.0 \times 10^7$ & 6400 & 0.8 & 340 & 425 & 0.020 & $1.70 \times 10^{-3}$ & 3.22 & M \\
 F1 & 80 & $3.0 \times 10^7$ & 3000 & 0.5 & 242 & 484 & 0.012 & $1.21 \times 10^{-3}$ & 2.76 & M \\
 F2 & 80 & $6.0 \times 10^7$ & 6000 & 0.5 & 312 & 624 & 0.030 & $1.56 \times 10^{-3}$ & 2.86 & M \\
 F3 & 80 & $8.0 \times 10^7$ & 8000 & 0.5 & 345 & 690 & 0.018 & $1.73 \times 10^{-3}$ & 2.61 & M \\
 F4 & 80 & $10.0 \times 10^7$ & 10000 & 0.5 & 375 & 750 & 0.041 & $1.87 \times 10^{-3}$ & 4.16 & M \\
\end{tabular}
\caption{Runs with fixed flux boundaries and no source, at $E=2 \times10^{-4}$ and $Pm=40$. Rayleigh number $Ra$ and modified Rayleigh number $\cal{R}$, Prandtl number $Pr$ and Roberts number $q$ are shown. $Rm$ is the magnetic Reynolds number, $Dip$ is the dipolarity, $Ro$ is the Rossby number and $\tau$ is the number of diffusion times for the whole run. Reversal: N means no reversal, $M$ means a multipolar dynamo with many reversals but no Earth-like fields, E means a reversing dynamo with Earth-like magnetic fields for some time intervals. All runs in this table used a resolution of $Nr \times Nl \times Nm =$ 160 $\times$ 128 $\times$ 128. }
\label{table:cases1}
\end{center}
\end{table*}

\begin{table*}
 \begin{center}
\def~{\hphantom{0}}

\begin{tabular}{ccccccccccc}
 Run  & $Pr$ & $Ra$ & $\cal{R}$ & $q$ & $Rm$ & $Pe$ & $Dip$  & $Ro$ & $\tau$ & Reversal\\
 
\hline
 G1 & 40 & $6.0 \times 10^7$ & 6000 & 1 & 316 & 316 & 0.123 & $0.79 \times 10^{-3}$ & 1.96 & N \\
 H1 & 50 & $7.0 \times 10^7$ & 5600 & 0.8 & 315 & 394 & 0.116 & $0.79 \times 10^{-3}$ & 4.08 & N \\
 H2 & 50 & $1.0 \times 10^8$ & 8000 & 0.8 & 389 & 486 & 0.085 & $0.97 \times 10^{-3}$ & 2.11 & N \\
 H3 & 50 & $1.2 \times 10^8$ & 9600 & 0.8 & 421 & 526 & 0.081 & $1.05 \times 10^{-3}$ & 2.49 & N \\
 H4 & 50 & $1.5 \times 10^8$ & 12000 & 0.8 & 466 & 582 & 0.073 & $1.16 \times 10^{-3}$ & 3.14 & N \\
 H5 & 50 & $2.5 \times 10^8$ & 20000 & 0.8 & 588 & 735 & 0.054 & $1.47 \times 10^{-3}$ & 1.94 & E \\
 I1 & 55 & $1.0 \times 10^8$ & 7272 & 0.727 & 370 & 509 & 0.087 & $0.92 \times 10^{-3}$ & 1.15 & N \\
 I2 & 55 & $1.5 \times 10^8$ & 10909 & 0.727 & 449 & 618 & 0.072 & $1.12 \times 10^{-3}$ & 5.04 & N \\
 I3 & 55 & $2.5 \times 10^8$ & 18182 & 0.727 & 562 & 773 & 0.056 & $1.41 \times 10^{-3}$ & 6.09 & E \\
 J1 & 60 & $5.0 \times 10^7$ & 3333 & 0.667 & 253 & 379 & 0.121 & $0.63 \times 10^{-3}$ & 1.74 & N \\
 J2 & 60 & $5.5 \times 10^7$ & 3667 & 0.667 & 258 & 387 & 0.129 & $0.64 \times 10^{-3}$ & 2.81 & N \\
 J3 & 60 & $5.75 \times 10^7$ & 3833 & 0.667 & 310 & 465 &0.029 & $0.77 \times 10^{-3}$ & 3.73 & M \\
 J4 & 60 & $6.0 \times 10^7$ & 4000 & 0.667 & 314 & 471 & 0.038 & $0.78 \times 10^{-3}$ & 4.81 & M \\
 J5 & 60 & $8.0 \times 10^7$ & 5333 & 0.667 & 354 & 531 & 0.031 & $0.88 \times 10^{-3}$ & 2.21 & M \\
 J6 & 60 & $1.0 \times 10^8$ & 6667 & 0.667 & 389 & 583 & 0.020 & $0.97\times 10^{-3}$ & 8.33 & M \\
 J7 & 60 & $1.5 \times 10^8$ & 10000 & 0.667 & 434 & 651 & 0.069 & $1.09 \times 10^{-3}$ & 6.56 & E \\
 J8 & 60 & $2.5 \times 10^8$ & 16667 & 0.667 & 542 & 813 & 0.056 & $1.35 \times 10^{-3}$ & 6.11 & E \\
 K1 & 70 & $1.5 \times 10^8$ & 8571 & 0.571 & 407 & 713 & 0.066 & $1.02 \times 10^{-3}$ & 6.01 & E \\
 K2 & 70 & $2.0 \times 10^8$ & 11429 & 0.571 & 461 & 807 & 0.058 & $1.15 \times 10^{-3}$ & 9.34 & E \\
 K3 & 70 & $3.0 \times 10^8$ & 17143 & 0.571 & 540 & 946 & 0.060 & $1.35 \times 10^{-3}$ & 6.72 & E \\
 L1 & 80 & $2.5 \times 10^8$ & 12500 & 0.5 & 468 & 936 & 0.066 & $1.17 \times 10^{-3}$ & 3.38 & N* \\
 L2 & 80 & $3.5 \times 10^8$ & 17500 & 0.5 & 537 & 1074 & 0.065 & $1.34 \times 10^{-3}$ & 2.84 & N \\
 \end{tabular}
\caption{Runs with fixed flux boundaries and no source, at $E=10^{-4}$ and $Pm=40$. Rayleigh number $Ra$ and modified Rayleigh number $\cal{R}$, Prandtl number $Pr$ and Roberts number $q$ are shown. $Rm$ is the magnetic Reynolds number, $Dip$ is the dipolarity, $Ro$ is the Rossby number and $\tau$ is the number of diffusion times for the whole run. Reversal: N means no reversal, $M$ means a multipolar dynamo with many reversals but no Earth-like fields, E means a reversing dynamo with Earth-like magnetic fields for some time intervals. All runs in this table used a resolution of  $Nr \times Nl \times Nm =$ 160 $\times$ 128 $\times$ 128.}
\label{table:cases2}
\end{center}
\end{table*}

\begin{table*}
 \begin{center}
\def~{\hphantom{0}}
\begin{tabular}{cccccccccccc}
 Run  & $E$                  & $Pr$ & $Pm$ & $Ra$               & $Rm$ &  $Pe$ & $Ro$ 
      & q   & ${\cal R}$ & $\tau$ & Reversal \\
\hline
 CCB0 & $2.5 \times 10^{-5}$ &  1  & 2.5 & $6.75 \times 10^8$ & 504 & 202 &0.0040 
      & 2.5 & $4.219 \times 10^4$ & 2.75 & N \\
 CCB1 & $7 \times 10^{-5}$   & 40 & 20 & $3.0 \times 10^9$  & 766   & 1532 &0.0027  
      & 0.5 & $1.05 \times 10^5$  & 4.93 & E \\   
 CCB2 & $2 \times 10^{-4}$ &  114  & 57 & $1.25 \times 10^9$ & 686 & 1372 &0.0024 
      & 0.5 & $1.25 \times 10^5$ & 1.57 & M \\
 CCB3 & $7 \times 10^{-5}$   & 40 & 20 & $1.4 \times 10^9$  & 545   & 1090 & 0.0019  
      & 0.5 & $4.9 \times 10^4$  & 7.08 & N \\ 
 CCB4 & $7 \times 10^{-5}$   & 35 & 20 & $2.0 \times 10^9$  & 683   & 1195 & 0.0027  
      & 0.5714 & $8 \times 10^4$  & 4.84 & E \\      
 \end{tabular}
 
\caption{Runs with compositional convection boundaries. Rayleigh number $Ra$ and modified Rayleigh number $\cal{R}$, Prandtl number $Pr$ and Roberts number $q$ are shown. $Rm$ is the magnetic Reynolds number, $Dip$ is the dipolarity, $Ro$ is the Rossby number and $\tau$ is the number of diffusion times for the whole run. Reversal: N means no reversal, $M$ means a multipolar dynamo with many reversals but no Earth-like fields, E means a reversing dynamo with Earth-like magnetic fields for some time intervals. All runs used a resolution of $Nr \times Nl \times Nm =$ 160 $\times$ 192 $\times$ 192, except CCB4, where a resolution of 160 $\times$ 192 $\times$ 128 was used. }
\label{table:cases3}
\end{center}
\end{table*}

\newpage
\bibliography{reversals}

\end{document}